\pdfoutput=1

\documentclass[11pt]{article}
\usepackage{jheppub}

\usepackage{amsmath}
\usepackage{latexsym,amssymb}
\usepackage{graphicx,color,slashed}
\usepackage{bbm} 
\usepackage{mathtools}
\usepackage{ifpdf}

\newcommand{\cF}{\mathcal{F}}
\newcommand{\cG}{\mathcal{G}}

\newcommand{\cA}{\mathcal{A}}
\newcommand{\cB}{\mathcal{B}}
\newcommand{\cC}{\mathcal{C}}

\newcommand{\cH}{\mathcal{H}}
\newcommand{\cL}{\mathcal{L}}

\newcommand{\cN}{\mathcal{N}}
\newcommand{\cO}{\mathcal{O}}

\newcommand{\be}{\begin{equation}}
\newcommand{\ee}{\end{equation}}
\newcommand{\ba}{\begin{eqnarray}}
\newcommand{\ea}{\end{eqnarray}}
\newcommand{\nn}{\nonumber}

\newcommand{\N}{\mathcal{N}}

\def\E{{$E_{7(7)}$}}

\newcommand{\rf}[1]{(\ref{#1})}
\newcommand{\bea}{\begin{eqnarray}}
\newcommand{\eea}{\end{eqnarray}}

\def\bfzero{\relax{\rm I\kern-.18em 0}}
\def\bfone{\relax{\rm 1\kern-.35em 1}}
\def\twomat#1#2#3#4{\left(\begin{array}{cc}
\end{array}
\right)}

\newcommand{\La}{\Lambda}
\newcommand{\Si}{\Sigma}

\newcommand{\la}{\lambda}

\def\GB{{{\rm E}_4}}
\def\cC{{\cal C}}
\def\si{\sigma}

\title{\rm{\bf 
The Action with Manifest E7 Type Symmetry }}
\author{
\rm{\bf Renata Kallosh}}
\affiliation{Stanford Institute for Theoretical Physics and Department of Physics, Stanford University, Stanford, CA 94305, USA}

\abstract{ We generalize Cremmer-Julia 1st order action of $\cN=8$ supergravity with  manifest \E\, symmetry to cases of $\cN=6$  with manifest $SO^*(12)$  and $\cN=5$ with manifest $SU(1,5)$ duality symmetries. These U dualities belong to groups of type E7 which do not  admit a symmetric  bilinear 
 invariant for vector fields.  Therefore the 2d order classical action derived from the one with manifest  E7 type duality  has a ghost vector field which,  under appropriate boundary conditions,  decouples.
 We show that when classical $\cN\geq 5$ supergravities are deformed by a candidate UV divergence the ghost field does not decouple. Therefore we argue that U duality and supersymmetry suggest an explanation of the  mysterious  cancellation of  UV infinities at $L=4$, $\cN=5$ in d=4.   
The same reasoning implies that, in absence of 
 duality and supersymmetry anomalies, which still require a better understanding, $\cN\geq 5$ perturbative supergravities may be UV finite at  higher-loops.}

\begin{document}

\maketitle



\parskip 5pt


\section{Introduction}

The \E\,  duality symmetry in $\cN=8$ supergravity  \cite{Cremmer:1978ds,Cremmer:1979up,deWit:1982bul} was discovered   by  Cremmer and Julia.   In general case of $\cN\geq 3$ extended supergravities the scalars are coordinates of the ${\cG\over \cH}$ coset space, where $\cG$ is a U duality group and $\cH$ is an isotropy. Duality symmetries  in nonlinear electrodynamics and extended supergravity were studied in  \cite{Gaillard:1981rj,Aschieri:2008ns}.

In ${\cal N}=5, 6, 8$ supergravities the relevant duality groups $\cG$ are:~$SU(1,5), \, SO^*(12), \, E_{7(7)}$. They are known as 
  {\it groups of type} E7  \cite{Brown,Garibaldi,Borsten:2011nq,Ferrara:2011dz,Ferrara:2012qp,Borsten:2018djw}. We review groups of type E7,  in particular with regard to $\cG$-duality in supergravity, in Appendix A.    The  main property of these groups relevant to our analysis here is that the vector field representation is symplectic and there are no bilinear symmetric invariants.\footnote{For $\cN\leq 3$ $\cG$-duality  groups might be degenerate, when the quartic invariant ÒdegeneratesÓ into a perfect square. But these models have $\cG$-duality anomalies anyway,  as shown in \cite{Marcus:1985yy}, and are not expected to be fully protected by $\cG$-duality against UV divergences.} The isotropy groups  $\cH$ of the  ${\cG\over \cH}$ coset space  are:~$U(5), U(6), SU(8)$  for  ${\cal N}=5, 6, 8$ supergravities, respectively.

  A proposal to study UV properties of supergravities  via their consistent deformation by the candidate UV divergences was made  by Bossard and Nicolai (BN) in \cite{Bossard:2011ij}, following the observation in \cite{Kallosh:2011qt,Kallosh:2011dp} that adding a UV divergence  candidate to  the classical action breaks  duality current conservation. The deformation we are discussing here may be viewed as an example of the following  procedure. We add a $\lambda  \phi^4$ interaction to Yukawa theory $ g \bar \psi \gamma^5 \psi \phi$ with the purpose to create a  deformed classical action, with the deformation parameter $\la$. This new deformed  action is now capable of absorbing the UV infinities of the loop computations, as opposite to the original one.

  More recently 
  ${\cal N}\geq 5$ supergravities, and their deformation due to candidate UV divergences, were studied  in \cite{Kallosh:2018mlw}.  A significant progress in studies of extended supergravities was achieved there, by making the analysis of duality symmetry universal for all $\cN\geq 5$ supergravities, based on symplectic section formalism \cite{Andrianopoli:1996ve}. These theories are anomaly free\footnote{We discuss the current status of anomalies in supergravities  in Appendix B.}  as shown in  \cite{Marcus:1985yy,Freedman:2017zgq} in the context of the 2d order formalism, and therefore one  expects that off-shell symmetries control quantum corrections.\footnote{  $\cN\leq 4$ supergravities interacting with matter are discussed in  Appendix C.} However, an attempt to extract an information about  UV infinities in $\cN\geq 5$ in  \cite{Kallosh:2018mlw}, based on  a BN deformation proposal \cite{Bossard:2011ij} in the 
2d order action,  was inconclusive.

The Lorentz covariant candidate counterterms in $\cN=8$ theory were constructed in \cite{Kallosh:1980fi,Howe:1980th,Howe:1981xy}. 
 A part of duality symmetry, based on soft limit on scalars, have been shown to explain some of the properties of UV finite amplitudes in extended supergravities. For example, in \cite{Beisert:2010jx} it was argued that $\cN=8$ supergravity is protected from UV divergences in d=4 up to 6 loop order, based on supersymmetry and \E\, symmetry. This explains the computations in \cite{Bern:2007hh,Bern:2009kd} which have shown UV finiteness at 3 and 4 loops.  But the prediction of \cite{Beisert:2010jx} is still to be confirmed at 5 and 6 loops in $\cN=8$ d=4. The analysis in \cite{Beisert:2010jx}, based on {\it soft limit on scalars} due to \E\, symmetry, is inconclusive  starting from 7 loops.

Meanwhile for $\cN=8$ other arguments were given about all-loop finiteness based on the light-cone formalism \cite{Kallosh:2010kk}. It is interesting that the light-cone finiteness argument in \cite{Kallosh:2010kk} was not disputed, but  the light-cone candidate counterterms at loop order $L$ were not constructed either. Therefore they are not known to exists in $\cN=8$ supergravity. A consistent supersymmetric reduction to $\cN=5,6$ was not studied so far.

A different argument about finiteness of $\cN=8$ supergravity is based 
 on  \E\, symmetry in the {\it vector sector} of the theory \cite{Kallosh:2011qt,Kallosh:2011dp}. The  \E\, symmetry  argument in \cite{Kallosh:2011qt,Kallosh:2011dp} was disputed by BN in \cite{Bossard:2011ij},  where a proposal was made that  the \E\, symmetry can be restored even in presence of the candidate counterterms. The proposal is based on a construction of the source of deformation, defined in details in \cite{Carrasco:2011jv}. The source of deformation has a  manifest \E\, symmetry, where instead of a physical vector in representation $\mathbf{28, \overline {28}}$ of $SU(8)$ one has to use a symplectic doublet with twice as many vector fields. An improved version of this proposal for $U(1)$ duality group was developed in 
\cite{Carrasco:2011jv} and it was applied to  the Born-Infeld theory, as well as Born-Infeld theory with higher derivatives \cite{Chemissany:2011yv,Chemissany:2006qd}.
The proof of  consistency of the deformation proposal in bosonic theory was given in \cite{Bossard:2011ij} only at a base point of the moduli space, where all scalar fields vanish.

The compatibility of the BN deformation proposal with supersymmetry was questioned in  \cite{Kallosh:2012yy} and in \cite{Gunaydin:2013pma} and  obstructions to this deformation were pointed out.
The actual computations in d=4 $\cN=8$ at 7-loop level, which would resolve the issues above,  are far too difficult, and results are not expected anytime soon.

However, for $\cN=5$ it became known about four years ago that  UV divergences are absent at  3 and 4 loop level \cite{Bern:2014sna}. Until recently there was no explanation of these computations. The current situation is the following. The soft scalar limit analysis in  \cite{Beisert:2010jx} for $\cN=8$ was generalized for the case of $\cN=5,6$ in \cite{Freedman:2018mrv}. The result is that consistency of the soft limit on scalars of amplitudes  with duality and supersymmetry  for $\cN\geq 5$  requires that at the loop order $L=\cN-2$ the theory is protected from UV divergences.  Thus, $\cN=5$ has to be UV finite at 3 loops, which explains part of the computation in \cite{Bern:2014sna}.
$\cN=6$ has to be UV finite at 4 loops,  $\cN=8$ has to be UV finite at 6 loops, which are  predictions still to be validated. The case of a critical loop order
\be
L_{cr}=\cN-1\ee
for all these theories, $\cN=5$  at 4 loops, $\cN=6$  at 5 loops, $\cN=8$  at 7 loops, remains elusive
 when only the soft limit analysis of amplitudes following from duality is combined with supersymmetry. A harmonic superspace analysis of available supersymmetric and duality invariant counterterms was performed in \cite{Bossard:2011tq}, where  UV divergence was predicted at $L_{cr}=\cN-1$ for $\cN\geq 4$.

Thus, not a single explanation of the $\cN=5$,  $L_{cr}=\cN-1= 4$ UV finiteness discovered four  years ago in \cite{Bern:2014sna} is available at present.  Why UV infinities in 82 diagrams cancel? We show the corresponding set of diagrams  in the Appendix D. 

The first hint  of a crisis with the 2d order deformed action in the  the BN approach  \cite{Bossard:2011ij}
 is the fact that  the proof of the  consistency of the deformation in supergravity with scalars is still missing now, eight years later. We explain  in our Appendices E and F why it was not yet possible to extend the BN proof in \cite{Bossard:2011ij}, which was made for vanishing scalars, to the case of the full theory where scalars are present. The second hint is in \cite{Kallosh:2012yy,Gunaydin:2013pma}, where the supersymmetry obstructions to  the deformation proposal  in \cite{Bossard:2011ij} were exposed.

Here we start a new direction of investigation using the 1st order  formalism  \cite{Cremmer:1979up} with manifest \E\, symmetry as a starting point. We generalize it
 to $\cN\geq 5$ supergravities with their relevant dualities.
 The most important for our purpose property of theories with E7 type symmetries is the absence of a bilinear  symmetric
 invariant. This is a fundamental reason why one encounters bad  ghosts when one tries to develop the BN deformation.
   In classical theory these bad ghosts decouple, as shown in   \cite{Cremmer:1979up}. However, when the 4-point vector deformations due to candidate UV divergences are included, bad ghosts do not decouple, as we will show below.  
 We will conclude that  the 2d order deformed  theory,   which follows from the 1st order one, with off-shell E7 symmetry, is inconsistent since it has ghosts. Meanwhile, when one starts with the conventional 2d order theory without ghosts, the proof of E7 symmetry of the deformed 2d order theory in  \cite{Bossard:2011ij} cannot be extended in presence of scalars.

Thus our main claim is that that E7 symmetry in $\cN\geq 5$ supergravities is inconsistent with the 4-vector UV divergence.  When we assume in addition an  unbroken $\cN\geq 5$ supersymmetry, which requires all other 4-point UV divergences to show up in computations with the same factor as a 4-vector one, we conclude  that our new observation suggests an explanation of the UV finiteness of $\cN=5$ at $L=4$ \cite{Bern:2014sna}, which was not explained so far. Moreover, our claim about the inconsistency of the 4-vector UV divergence in $\cN\geq 5$ supergravities with E7 symmetry is valid, conditionally,  at any loop order. The necessary condition is that  duality and supersymmetry anomalies are absent. 
Thus our observation suggests that in the absence of  duality and supersymmetry anomalies $\cN\geq 5$ supergravities may be UV finite.

 \section{The manifestly $\cG$-duality  invariant vector action for $\mathcal{N}\geq 5$ supergravity} 
 
We explain here the important  technical features of the symplectic section formalism, developed in \cite{Andrianopoli:1996ve} and used in the studies of UV infinities of perturbative supergravity in
 \cite{Kallosh:2012yy} and, more recently, in \cite{Kallosh:2018mlw}.

 In $\cN=8,6,5$ the number of physical vectors is $n_v= (28, 16,10)$ respectively. However, the manifest E7 type symmetry in these models 
requires that the action depends on duality doublets, which have twice an amount of vectors: $n_{2v}= (56, 32,20)$. Therefore we will need also another vector duality doublet in the action, as well as a Lagrange multiplier to a duality invariant constraint and a doublet depending on scalars, to be able to construct an action, classical or deformed,  with a manifest E7 type symmetry.

\subsection{Bilinear symplectic invariants and graviphotons }

\label{sec-graviphotons}
Consider  a $n_{2v}$-dimensional real symplectic
vector of field strengths
\begin{equation}\label{calF}
\mathcal{F} \equiv 
\left(
  \begin{array}{c}
  F^{\Lambda} \\ G_{\Lambda} \\  
  \end{array}
\right)\, .
\end{equation}
that transforms in the $\mathbf{56,32,20}$ of the corresponding $\cG$-duality groups. A doublet transforms as follows
\begin{equation}
\left ( \begin{array}{c}  F\cr  G\cr   \end{array}\right
)^\prime \, =\, \left ( \begin{array}{cc} A & B \cr C & D \cr  \end{array} \right )
\left (  \begin{array}{c}   F\cr   G\cr  \end{array}\right ).
\label{dualrot}
\end{equation}
Here a constant matrix ${\cal S} =\left (\begin{array} {cc}A & B \cr C & D
\cr  \end{array} \right ) \, \in \, GL(2{ n_v},\mathbb{R} )$.
The scalars of the theory are described by the symplectic section
\begin{equation}
\label{eq:symplecticsection}
\mathcal{V}_{AB}
\equiv
\left(
  \begin{array}{c}
  f^{\Lambda}{}_{AB} \\ h_{\Lambda\, AB} \\  
  \end{array}
\right)\, ,
\end{equation}
where $A,B=1,\cdots,\cN$ are an antisymmetric pair of indices for the $\cH$ isotropy that are raised
and lowered by complex conjugation. The period matrix is defined as follows
$
h_{\Lambda\, AB} = \mathcal{N}_{\Lambda\Sigma}f^{\Sigma}{}_{AB}$.
For the symplectic product $\langle ~ \mid ~ \rangle$, we
  use the convention
\begin{equation}
\langle  \mathcal{A}\mid \mathcal{B}\rangle \equiv
  \mathcal{B}^{\Lambda}\mathcal{A}_{\Lambda} 
-\mathcal{B}_{\Lambda}\mathcal{A}^{\Lambda}\, .
\end{equation}
The graviphoton field strength is defined by
\begin{equation}
\label{eq:graviphotondef}
T_{AB}  
\equiv
\langle \mathcal{V}_{AB}\mid \mathcal{F} \rangle\, , 
\end{equation}
and its self- and anti-selfdual parts are
\be
T_{AB}{}^{\pm}  
\equiv
\langle \mathcal{V}_{AB}\mid \mathcal{F}{}^{\pm}   \rangle\, , \qquad   {T}^{* \pm AB}  
\equiv
\langle \overline {\mathcal{V}}^{AB}\mid \mathcal{F}{}^{\pm}   \rangle \ .
\ee
Graviphotons are $\cG$-duality invariant, they transform under compensating $(S)U(\cN)$ transformations only. In classical $\N\geq 5$ supergravity in absence of fermions there is a  linear twisted self-duality constraint %
\begin{equation}
\label{linear}
{T}_{AB}{}^{ +} =  h_{\Lambda
 AB}\, F^{+\Lambda}_{\mu\nu} - f^\Lambda_{AB} \,G_{ \mu\nu\,\Lambda}^+ =0 \ ,  \qquad   {T}^{*- AB} =\bar h_{\Lambda}^{
 AB}\, F^{-\Lambda}_{\mu\nu} - \bar f^{\Lambda AB} \,G_{ \mu\nu\,\Lambda}^- =0\, . 
\end{equation}
It results in the relation between $G$ and $F$, so that only one of them is independent
\be
G^+= \cN F^+\, , \qquad G^-= \overline \cN F^- \, .
\ee 
This gives a correct amount of the physical degrees of freedom for vectors, $\mathbf{28,16,10}$  and is  one-half of the symplectic representation of the $\cG$-duality symmetry  for \E\,, $SO^*(12)$ and $SU(1,5)$ duality, respectively.

\subsection{Vectors and fermions in classical theory in the  2d order form action}
In classical supergravity adding fermions in the context of duality requires to add to the bosonic part of the action which is quadratic in $F$, also  a term linear in $F$, which is scalar dependent and quadratic in fermions as well as a scalar dependent term quartic in fermions.
The $\cH$-covariant  combination of fermions $\cO _{\mu\nu  AB} ^{+ }$, $\cO _{\mu\nu  } ^{- AB}$ has terms with products of two  spin 1/2 fields, two gravitino's and a spin 1/2 and a gravitino, shown for example in eq. (7) in \cite{Cremmer:1978ds} and in 
eq. (2.22) in \cite{deWit:1982bul} in $\cN=8$ theory.

We present the relevant part of the classical $\cN\geq 5$ supergravity action in terms of a symplectic section formalism \cite{Andrianopoli:1996ve} which was used recently in \cite{Kallosh:2018mlw} in the bosonic theory without fermions, here we also include fermions. Our discussion is universal for $\cN=5,6,8$.
\be
S =
  F\tilde G -  i \, T^-_{ AB} \cO^{-AB}  + i \, T^{*+ AB} \cO^+_{AB} \label{actionGZcl1} \ .
\ee
The modified constraint in presence of fermions relating $G$ to $F$ and scalars and fermions is
\be
 T^{+}_{\mu\nu\, AB} \equiv   h_{\Lambda
 AB}\, F^{+\Lambda}_{\mu\nu} - f^\Lambda_{AB} \,G_{ \mu\nu\,\Lambda}^+  =i\, 
 \cO^+_{AB} \ ,
\ee
which means that 
\be
G_{ \mu\nu\,\Lambda}^{+}= -i\, f^{-1 AB}_{ \La}   \cO^+_{AB}+ \cN_{\La \Si} F^{+\Si} \ ,
\label{G0}\ee
i. e. $G$ depends on $F$, on scalars and on fermions. 
We can also present the action \rf{actionGZcl1} as
\be
S =
i F^- G^- - i F^+ G^+ -  iT^-_{ AB} \cO^{-AB}  +i T^{*+ AB} \cO^+_{AB} \label{ac} \ ,
\ee
since
$
 \tilde G = i(G^-  -  G^+) 
$.
We define the constraint in presence of fermions as 
\be
 T^{+\,  \cO}_{\mu\nu\, AB} \equiv   h_{\Lambda
 AB}\, F^{+\Lambda}_{\mu\nu} - f^\Lambda_{AB} \,G_{ \mu\nu\,\Lambda}^+  -i\,   \cO^+_{AB}=0 \ .
\label{O}\ee
Now we prepared the tools we need to present manifestly E7 type symmetric actions for classical  $\cN\geq 5$ supergravities.

\subsection{The 1st order action} 

The   action in the 1st order formalism  depending on vectors is manifestly invariant under $\cG$-duality, i.e. E7 type symmetry, as well as under  $\cH$ isotropy group,  $SU(8), U(6), U(5)$ for $\cN=8,6,5$ respectively.
\be
 \cL^{1st} =  -   \langle \,  \cF_1  \,  |  \, \tilde  \cF_2\,  \rangle  +i \, T^{*+ AB}_2 \cO^+_{AB} +i \,  L^{AB+}  {T}_{2 \, AB }{}^{ + \, \cO} +h.c.  
\label{Action}\ee
This action is the same, in different notations,  as the 1st order formalism action presented for $\cN=8$ in \cite{Cremmer:1978ds,Cremmer:1979up}.
Here
\begin{equation}
-\langle  \cF_1\mid \tilde  \cF_2\rangle \equiv
  -\tilde  F_2^{\Lambda} G_{1\Lambda} 
+\tilde G_{2\Lambda}F^{\Lambda}_1=    i F_2^+ G_1^+ - i G_2^+ F_1^+ +h.c.
\end{equation}
The manifestly E7 type and $\cH$-invariant  action  \rf{Action} depends on 2 independent vector symplectic doublets.
The first doublet is a field strength  
\be
 \cF_1= d \cA_1 \qquad {\rm \, off-shell}
 \ee 
with the doublet 1-form vector potential $\cA_1$:
\begin{equation}
\cF_{1\mu\nu} = \partial_{[\mu } \cA_{1\nu ]} =\left ( \begin{array}{c}   F_1^\La \cr   G_{1 \La}\cr   \end{array}\right )_{\mu\nu}=\left ( \begin{array}{c}    \partial_{[\mu } \cB_{1\nu ]} \cr    \partial_{[\mu } \cC_{1\nu ]} \cr   \end{array}\right ).
\label{cF1}
\end{equation}
 The second doublet is an antisymmetric tensor (off shell it is not a field strength)
\begin{equation}
\cF_{2\mu\nu}  =\left ( \begin{array}{c}   F_2^\La \cr   G_{2 \La } \cr   \end{array}\right )_{\mu\nu} \ .
\label{cF2}
\end{equation}
The Lagrange multiplier $L^{AB+}$ and a graviphoton constraint ${T}_{AB}{}^{ + \, \cO }$ in eq. \rf{O}, and their conjugates, are E7-duality invariants, they transform under $\cH$. 
The symbol ${T}_{2 \, AB }{}^{ + \cO}$ in the action means that it depends on the vector doublet $\cF_2$ and on fermions
\begin{equation}
T_{2 \, AB }{}^{+ \cO}  
\equiv
\langle \mathcal{V}_{AB}\mid \mathcal{F}{}^{+}_2   \rangle\,  -i  \cO^+_{AB} \ .
\end{equation}
 {\it Off shell we treat $\cF_1$ and $\cF_2$ in an asymmetric way}.  This has an advantage that the equation of motion over $\cA_1$ will produce the requirement that  on shell
$
d\cF_2=0
$. For this to happen, it is necessary that 
 $\cF_1$ appears in the action only once, in a bilinear invariant formed with two doublets
\be
-\int d^4 x  \, \langle \,  \cF_1  \,  |  \, \tilde  \cF_2\,  \rangle= -\int d^4 x  \, \langle \,  d\cA_{1\mu}  \,  |  \, \tilde  \cF_2\,  \rangle \ .
\label{action1}\ee
Meanwhile, $\cF_2$ appears in the first term,  in the second term in the interaction with fermions,  and also in the third term with a Lagrange multiplier in our action \rf{Action}.

\section{$\cG$-duality  covariant equations of motion}
After partial integration one finds that the  vector potential $\cA_{1\mu} $ is a Lagrange multiplier to a doublet field equation 
 \be\label{logic}
{\delta S\over \delta \cA_{1\mu}}=0 \quad \Rightarrow \quad \partial_\nu  \tilde  \cF_2^{\mu\nu} \approx 0 \, .
\ee
Therefore on shell 
\be\label{logic1}
 \cF_2 \approx d\cA_2\,    \qquad {\rm \, on-shell}
\ee
i.e. on shell there is a second doublet vector potential $\cA_{\mu 2}$ and $\cF_{2\mu\nu} \approx \partial_{[\mu } \cA_{2\nu ]}$. 
\begin{equation}
\cF_{2\mu\nu} \approx \partial_{[\mu } \cA_{2\nu ]} =\left ( \begin{array}{c}   F_2^\La \cr   G_{2 \La}\cr   \end{array}\right )_{\mu\nu}\approx \left ( \begin{array}{c}    \partial_{[\mu } \cB_{2\nu ]} \cr    \partial_{[\mu } \cC_{2\nu ]} \cr   \end{array}\right )\, .
\label{cF2n}
\end{equation}
We now differentiate the action \rf{Action}  over the Lagrange multiplier   and we find that
\be
{\delta S\over \delta L^{AB+}}=0 \quad \Rightarrow \quad  {T}_{2 \, AB }{}^{ + \,\cO }\equiv  h_{\Lambda
 AB}\, F^{+\Lambda}_{2 \, \mu\nu } - f^\Lambda_{AB} \,G_{ 2 \, \mu\nu\,\Lambda }^+  -i\,  \cO^+_{AB} \approx 0\, .  \label{Con} \ee
Next equation of motion is over $\cF_2$. The relevant terms in the action are
\be
iF_2^+ G_1^+ - iG_2^+ F_1^+ +  i  \cO^+ (\bar h F_2^+ - \bar f G_2^+)   + i L^+ ( hF_2^+ - fG_2^+ -i\cO^+) \, .
\ee
We differentiate the action over $F_2^{+ \Si}$ and $G_{2 \Si}^{+ }$ and find
\be
{\partial \cL \over \partial F_2^{+ \Si}} =0 \quad \Rightarrow \quad   G_{1\Si}^+ + \cO ^{ +}_{AB} \bar h_\Si ^{AB}  + L^{AB +} h_{AB \Si}  \approx  0\, ,
\label{1c}\ee
\be
{\partial \cL \over \partial G_{2 \La}^{+ }}= 0 \quad \Rightarrow \quad  F_{1}^{\La +} + \cO ^{ +}_{AB} \bar f ^{AB\La}+     L^{AB +}f_{AB}^\La \approx 0\, .
\label{2c}\ee
 We solve these equations eliminating the Lagrange multiplier and find that
\be
G_{1\La}^+ \approx   \cN_{\La \Si} F_{1}^{\Si +} -i f^{-1 AB}_\La  \cO^+_{AB}\, \quad \Rightarrow \quad   {T}_{AB \, 1}{}^{ + \cO}\approx 0 \ .
\ee
Compare this with the eq. \rf{G0} which solves the constraint \rf{O} applied to $\cF_2$,  i. e. when ${T}_{AB \, 2}{}^{ + \cO}\approx0$ so that 
\be
G_{2\La}^+ \approx   \cN_{\La \Si} F_{2}^{\Si +} -i f^{-1 AB}_\La  \cO^+_{AB}\, \quad \Rightarrow \quad   {T}_{AB \, 2}{}^{ + \cO}\approx 0 \ .
\ee
Thus we see that on shell there is a complete symmetry between the first and the second E7 doublet
\be
 \cF_{1\mu\nu} = \left ( \begin{array}{c}    \partial_{[\mu } \cB_{1\nu ]} \cr    \partial_{[\mu } \cC_{1\nu ]} \cr   \end{array}\right )\, , 
 \quad \cF_{2\mu\nu} \approx \left ( \begin{array}{c}    \partial_{[\mu } \cB_{2\nu ]} \cr    \partial_{[\mu } \cC_{2\nu ]} \cr   \end{array}\right ),
\ee
and 
\be
{T}_{1 \, AB }{}^{ + \cO}\approx 0 \, , \qquad {T}_{2 \, AB }{}^{ + \cO} \approx 0 \ .
\label{sym}\ee
All equations of motion are manifestly E7 type covariant! The symmetry between $ \cF_{1\mu\nu}$ and $ \cF_{2\mu\nu}$ is restored on shell.

Note that with $G_2$ and $G_1$ depending on $F_2$ and $F_1$ we still have twice the number of fields versus physical vectors. Namely, with independent  $  \mathbf{28,16,10} $ for $\cB_{1\nu } $ and  $  \mathbf{28,16,10}$ for $\cB_{2\nu } $ we still have a double set of fields,  in $\cN=8,6,5$ respectively. We will see how in the 2d order formalism the field $\cB_{1\nu } + \cB_{2\nu } $ will become a physical field, 
\be
\cB_{1\nu } + \cB_{2\nu } \equiv 2B_\nu
\ee
whereas $\cB_{1\nu } - \cB_{2\nu } $ will become a ghost field with a wrong sign of the kinetic term
\be
\cB_{1\nu } - \cB_{2\nu } \equiv 2 \mathbb{B}_\nu
\ee

\section{From the 1st to the 2d order action: ghosts decouple}

At the classical 2d order theory  half of the E7 type symplectic doublets corresponds to physical vectors,  $\mathbf{28,16,10}$ in $\cN=8,6,5$ respectively. There are  $\mathbf{28,16,10}$ equations of motions as well as  $\mathbf{28,16,10}$ Bianchi identities. The 
E7 type symmetry flips one into another. The corresponding E7 type symmetry is not manifest.

From the manifestly E7 type invariant 1st order action in \rf{Action} we proceed with derivation of the 2d order action, as follows.
\begin{itemize}
  \item We integrate over Lagrange multiplier, i. e.  we use the condition ${T}_{AB \, 2}{}^{ + \cO}=0$ which means that 
 \be
 G_{2\La}^+ \approx    \cN_{\La \Si} F_{2}^{\Si +} -i f^{-1 AB}_\La  \cO^+_{AB}\, .
\label{G2} \ee
 The remaining action 
\be \label{act}
 -\langle \,  \cF_1  \,  |  \, \tilde  \cF_2\,  \rangle  +i T^{*+ AB}_2 \cO^+_{AB}  +h.c. \ee
  depends on $F_{1 \mu\nu}= \partial_{[\mu } \cB_{1\nu ]}$,  $G_{1 \mu\nu}= \partial_{[\mu } \cC_{1\nu ]}$, on $F_2$,  as well as on scalars and fermions.
  \item We integrate the action over $\cC_{1 \mu}$ now. The only term depending on it is $ \cC_{1 \mu} \partial_\nu \tilde  F_2^{\mu\nu} $. This equation is solved if $F_{2 \mu\nu}\approx  \partial_{[\mu } \cB_{2\nu ]}$

\item Our 1st order action becomes
\begin{equation}
  - i G_2^+ F_1^+ +i T^{*+ AB}_2 \cO^+_{AB} +h.c.
 \end{equation}
 With account of the constraint on $G_2$ in \rf{G2} it becomes
  \begin{equation}
- i  F_1^{+\La} \cN_{\La \Si} F_{2}^{\Si +} -  (F_1^{+\La} + F_2^{+\La}) f^{-1 AB}_\La  \cO^+_{AB}  -   f^{-1 CD}_\La  \bar f^{\La AB}\cO^+_{CD} \cO^+_{AB} +h.c.
\label{action2} \end{equation}
or equivalently
  \begin{equation}
- i  \partial_{[\mu } \cB_{1\nu ]}^{+\La} \cN_{\La \Si} \partial^{[\mu } \cB_{2 ]}^{ \nu \Si +} -  \partial_{[\mu } (\cB_1+ \cB_ 2)_{\nu ]})^{+\La} f^{-1 AB}_\La  
\cO^{+ \mu\nu} _{AB} -   f^{-1 CD}_\La  \bar f^{\La AB}\cO^+_{CD} \cO^+_{AB} +h.c.
\label{action2} \end{equation}
\end{itemize}

\subsection{Why there are ghosts?}
The fundamental reason why we are facing ghosts in the 2d order formalism originating from the 1st one, is that the manifest E7 type symmetry can only operate with the number of fields twice as big as the  number of physical degrees of freedom. Technically, we see this as follows.
Let us look carefully at the kinetic term
  \begin{equation}
- i  F_1^{+\La} \cN_{\La \Si} F_{2}^{\Si +}  +h.c.
\label{kin} \end{equation}
and use notation
\be
F_1+F_2\equiv 2 F\, , \qquad F_1-F_2\equiv 2\mathbb{F} \ ,
\ee
so that 
$
F + \mathbb{F}=  F_1,\ F - \mathbb{F}=  F_2,
$ and 
  \begin{equation}
- i  (F + \mathbb{F}) ^{+\La} \cN_{\La \Si} ( F - \mathbb{F}) ^{\Si +}  +h.c. = -i  F ^{+\La} \cN_{\La \Si}  F ^{\Si +} +i  \mathbb{F} ^{+\La} \cN_{\La \Si}  \mathbb{F} ^{\Si +} +h.c.
 \end{equation}
The same can be presented as
 \begin{equation}
 -i  \partial_{[\mu } B_{\nu ]}
 ^{+\La} \cN_{\La \Si}  \partial_{[\mu } B_{\nu ]}
^{\Si +} +i  \partial_{[\mu } \mathbb{B}_{\nu ]}^{+\La} \cN_{\La \Si}  \partial_{[\mu } \mathbb{B}_{\nu ]} ^{\Si +} +h.c.
 \end{equation}
or as 
  \begin{equation}
 F  ^{\La} \tilde G_{\La} -  \mathbb{F} ^{\La}  \tilde {\mathbb{G}} _{\La } = -{1\over 4} F_{\mu\nu}^2 +{1\over 4} \mathbb{F}_{\mu\nu}^2
+ \cdots
\label{plusminus} \end{equation}
 Thus we see that the combination of the fields 
\be
F_1+F_2=  \partial_{[\mu } \cB_{1\nu ]} +  \partial_{[\mu } \cB_{2\nu ]}\equiv  2  \partial_{[\mu } B_{\nu ]}
\ee 
behaves as a normal vector field with the correct sign of the kinetic term, whereas the combination 
\be
F_1-F_2=  \partial_{[\mu } \cB_{1\nu ]} -  \partial_{[\mu } \cB_{2\nu ]}\equiv  2  \partial_{[\mu } \mathbb{B}_{2\nu ]}
\ee
 has a wrong sign of the kinetic term and is therefore qualified as a ghost vector field. In classical theory we will see that the ghost is decoupled, however, in a theory deformed by candidate counterterms, they do not decouple.
 
 If we would not be restricted by $\cG$-duality symmetry  that there are no symmetric bilinear invariants, we would be able to add additional terms to the action, 
and the situation would have to be reconsidered.

But for the non-degenerate groups of type E7, additional terms do not exist in the  1st order action, and kinetic terms for ghosts might cause a problem, depending on the interaction Lagrangian.

Let us explain the reason for the negative sign and ghosts in simple terms. We find the expression in \rf{kin} of the kind $a\cdot b$. There are no  terms $a\cdot a $ or $b\cdot b$ due to non-degenerate type E7 symmetry group. But we like to express $a\cdot b$  via a combination of some diagonal terms. So we define $a+b = 2x$ and $a-b=2y$ and find that
\be
a\cdot b= (x+y) \cdot (x-y) = x\cdot x - y\cdot y
\ee
 It is clear that these two diagonal terms must have opposite signs. This is why we get one sign for the physical field kinetic term in \rf{plusminus} and the opposite sign for the unphysical field kinetic term.  Wrong sign kinetic terms are sources of instability, therefore such fields are called ghosts and are not acceptable.

 \subsection{Why  do ghosts decouple in classical supergravity action?}
 
Our action \rf{action2} has the following dependence on all vector fields
 \be
 -i  F ^{+\La} \cN_{\La \Si}  F ^{\Si +} +i  \mathbb{F} ^{+\La} \cN_{\La \Si}  \mathbb{F} ^{\Si +}  -  2 F^{+\La}  f^{-1 AB}_\La  \cO^+_{AB} +h.c.
 \label{actionV}\ee
Only normal vectors interact with fermions, ghost vector fields do not interact with fermions. They do interact with scalars and gravity as one can see from the second term in eq. \rf{actionV}. In  \cite{Cremmer:1979up} the argument was given as to why the vector ghosts decouple. It was suggested to use classical equations of motion for the vector fields. For the normal vector one finds that on shell
\be
\partial_\mu \tilde G^{\mu\nu} \approx {\partial L_{int}^F \over \partial A_\nu} \ ,
\ee
where $L_{int}^F$ is a term where the normal vector interacts with fermions. One can check that the corresponding action does not vanish on shell with account of field equations. 

For the ghost vector field one finds, looking at eq. \rf{action2}, that  they are decoupled from fermions so that on shell
\be
\partial_\mu \tilde {\mathbb{G}} ^{\mu\nu} \approx 0 \ .
\ee
Here a boundary condition on $\cB_{1\nu } - \cB_{2\nu} = 2 {\mathbb{B}}_\nu=0$ at infinity is imposed consistently, as suggested in   \cite{Cremmer:1979up},  and the relevant part of the action, upon integration by part, vanishes on shell
\be
 \int   \mathbb{F} ^{\La}  \tilde {\mathbb{G}} _{\La } \approx \int {\mathbb{B}}_\nu \partial_\mu \tilde {\mathbb{G}} ^{\mu\nu} \approx 0 \ .
\ee
We are therefore left with the normal vector field and the 2d order action, including fermions, is
 \be
 -i  F ^{+\La} \cN_{\La \Si}  F ^{\Si +}   -  2 F^{+\La}  f^{-1 AB}_\La  \cO^+_{AB} -   f^{-1 CD}_\La  \bar f^{\La AB}\cO^+_{CD} \cO^+_{AB} +h.c.
  \label{actionV1}\ee
This is the standard 2d order $\cN\geq 5$ supergravity action for vectors and fermions coupled to scalars. We have generalized the $\cN=8$ setting in  \cite{Cremmer:1979up} to the case of $\cN=6,5$.

\section{Quartic  deformation due to candidate UV divergence}

The candidate counterterms in $\cN=8$ theory were constructed in \cite{Kallosh:1980fi,Howe:1981xy,Howe:1980th} and they are generalizable to $\cN=5,6$ via a consistent supersymmetric reduction. 
When the candidate UV divergences are added to the classical action, bosonic  linear twisted self-duality constraint 
\begin{equation}
\label{con}
{T}_{AB}{}^{ +  }\equiv   h_{\Lambda
 AB}\, F^{+\Lambda}_{\mu\nu} - f^\Lambda_{AB} \,G_{ \mu\nu\,\Lambda}^+ =0 \ ,
\end{equation}
is deformed following the proposal in  \cite{Bossard:2011ij}. The deformation of the constraint was proposed  in  \cite{Bossard:2011ij} starting with an expression $I( \cF)$ where $\cF$ is a vector $\cG$-duality doublet.   This   was improved in \cite{Carrasco:2011jv} where it was  explained that it must depend both on a vector duality doublet as well as on a scalar-dependent symplectic section, or alternatively on an $\cH$-covariant  graviphoton. This new expression $I (T^-,  T^{*+})$ was 
called in \cite{Carrasco:2011jv} a source of deformation. In  \cite{Bossard:2011ij} the deformed constraint is 
 $\cG$-covariant and has somewhat complicated dependence on derivatives of $I( \cF)$ over vector fields, involving the metric of the moduli space. In \cite{Carrasco:2011jv} 
 the deformed constraint is proposed in the $\cH$-covariant, $\cG$ invariant  form and the deformation is a derivative of the source of deformation over the graviphoton:
\begin{equation}
\label{conD}
{T}_{AB}{}^{ + \, def }\equiv  {T}_{AB}{}^{ +} - \la {\delta I (T^-,  T^{*+})\over \delta T^{*+AB}}=0 \ ,
\end{equation}
It was developed in details for $\cN\geq 5$ supergravities in
\cite{Kallosh:2018mlw}.  The solution of the constraint \rf{con} is known for a 2-graviton-2-vector source of deformation either in closed form or as a series in the deformation parameter $\la$, \cite{Kallosh:2018mlw}.
Since the deformation terms are complicated, we will first study bosonic actions without fermions. 

 The source of deformation we will use here is an example of a 3-loop candidate counterterm in $\cN=8$ theory \cite{Kallosh:1980fi,Howe:1981xy,Howe:1980th}. The bosonic part of it  is   \E\, invariant since it depends on 
\be
T_{AB}{}^{-}  
\equiv
\langle \mathcal{V}_{AB}\mid \mathcal{F}{}^{-}   \rangle\, , \qquad   {T}^{* + AB}  
\equiv
\langle \overline {\mathcal{V}}^{AB}\mid \mathcal{F}{}^{+}   \rangle \ .
\ee
 In  eq. \rf{tomas} below  we show the 4-vector part of it.  The difference between the candidate counterterm as presented in \cite{Kallosh:1980fi,Howe:1981xy,Howe:1980th}, and the source of deformation for the BN proposal in  \cite{Bossard:2011ij} or in \cite{Carrasco:2011jv} is, in particular,  that the counterterm is proven to be duality invariant and supersymmetric only when the classical constraint \rf{con} is used. However, the source of deformation must depends on unconstrained graviphotons, otherwise the variation over $T$ in eq. \rf{conD} would not be possible. Source of deformation has $F^\La$ and $G_\La$ as independent fields in \rf{calF} as defined in the graviphoton in \rf{eq:graviphotondef}. In the counterterm $G_\La$ is a functional of $F^\La$ and scalars.
 
 One can also insert some number of $SU(8)$ covariant derivatives in the 3-loop counterterm and in this way making it an L-loop candidate counterterm, since 
 by inserting derivatives we increase the dimension of the relevant counterterm and the corresponding  source of deformation represents a higher loop candidate counterterm. For example, starting with $L=8$ the relevant source of deformation is also manifestly supersymmetric, as well as  \E\, invariant when classical equations of motion are imposed. It is given by some superspace integral, symbolically
\be
\kappa^{14}\int d^4 x \, d^{32} \theta  \det E  (\chi \bar \chi )^2\, , 
\ee
where $\chi$ is a superfield representing a superspace torsion, \cite{Kallosh:1980fi,Howe:1980th}. The explicit  deformation related to a 3-loop candidate counterterm has less derivatives in the 4-point matrix elements,  \cite{Kallosh:1980fi,Howe:1981xy,Howe:1980th}. But as we will see below, changing the number of derivatives and the dimension of the deformation  plays no role in our analysis. 

\subsection{A solution of the deformed ``constraint'' for $G_2$}

Here the ``constraints" are actually equations of motion from the variation of the 1st order action over the Lagrange multiplier.
A quartic in vectors deformation  in the form given  in \cite{Kallosh:2012yy} is
\begin{equation}
\begin{array}{rcl}
I (T^{-},  T^{ *+})
& = & 
2\partial_{\mu}\partial_{\nu}T^{-}{}_{AB\, \alpha\beta}
T^{* +\, AB}{}_{\dot{\alpha}\dot{\beta}}
\partial^{\mu}T^{-}{}_{CD}{}^{\alpha\beta}
\partial^{\nu}T^{*+\, CD\, \dot{\alpha}\dot{\beta}}
\\
& & \\
& & 
+
\partial_{\mu}\partial_{\nu}T^{-}{}_{AB\, \alpha\beta}
\partial^{\mu}T^{*+\, BC}{}_{\dot{\alpha}\dot{\beta}}
T^{-}{}_{CD}{}^{\alpha\beta}
\partial^{\nu}T^{*+\, DA\, \dot{\alpha}\dot{\beta}}
+\mathrm{c.c.}
\end{array}
\label{tomas}\end{equation}
All derivatives can be easily extended to $(S)U(\cN)$ covariant ones with the help of the scalar dependent connections.
The required expression in \rf{con} ${\delta I (T^-,  T^{*+})\over \delta T^{*+AB}}$ is cubic in graviphotons $T$ and   has 8 contributions, starting with
\be
{\delta I (T^-,  T^{*+})\over \delta T^{*+AB}{}_{\dot{\alpha}\dot{\beta}}
}= 2\partial_{\mu}\partial_{\nu}T^{-}{}_{AB\, \alpha\beta}
\partial^{\mu}T^{-}{}_{CD}{}^{\alpha\beta}
\partial^{\nu}T^{*+\, CD\, \dot{\alpha}\dot{\beta}} +\cdots
\ee
In the approximation that we use the classical ``constraint" ${T}_{AB}{}^{ +  }=0$ ignoring the higher $\lambda$ terms one can replace
graviphotons as follows:
$
T_{AB} ^- \Rightarrow \bar f^{-1}_{AB \, \La} F^{-\La}$,   $T^{*+ AB }\Rightarrow f^{-1 AB}_\La F^{+ \La} $.
The ``constraint''  becomes
\be \label{con1}
{T}_{AB}{}^{ + \, def }\equiv  (h F)_{AB}{}^+ - (f G)_{AB}{}^+ - 2 \la   (\partial_{\mu}\partial_{\nu}\bar f^{-1} F^-{})_{AB}
(\partial^{\mu}\bar f^{-1} F^- {})_{CD}{}
(\partial^{\nu}f^{-1} F^{+  CD} )+\cdots =0 \ .
\end{equation}
We can solve it for $G$ in terms of $F$ and scalars and we find 
\be \label{sol}
 G^+_\La  = (\cN  F)_\La^+  - 2 \la f^{-1 AB}_\La (\partial_{\mu}\partial_{\nu}\bar f^{-1} F^-{})_{AB}
(\partial^{\mu}\bar f^{-1} F^- {})_{CD}{}
(\partial^{\nu}f^{-1} F^+)^ {CD}  +\cdots 
\end{equation}
 According to our strategy, we will look at the constraint on $\cF_2$ which will come with the Lagrange multiplier in the 1st order action
\be
 \cL^{1st} =  -   \langle \,  \cF_1  \,  |  \, \tilde  \cF_2\,  \rangle  +i \,  L^{AB+}  {T}_{2 \, AB }{}^{ + \, def } +h.c.
 \label{ActionDef}\ee
 Here we stress that there is no  arbitrariness in implementing the deformation
proposal in case of the non-degenerate duality groups of type E7: the Lagrange multiplier term with $L^{AB+}$  can only depend on $\cF_2$. If it would depend also on $\cF_1$, this would destroy the design of the construction, explained in eqs. \rf{logic}, \rf{logic1}. Namely 
 the  vector potential $\cA_{1\mu} $ is a Lagrange multiplier to a doublet field equation for $\cF_2$. The single dependence on $\cF_1$ must be  in the first term in eq. \rf{ActionDef}, so that on shell $ \partial_\nu \tilde  \cF_2^{\mu\nu} \approx 0$ and $\cF_{2\mu\nu} \approx \partial_{[\mu } \cA_{2\nu ]}$. Deformed expression for ${T}_{AB }^+$ can only depend on $\cF_2$, one cannot split the deformation between the two vector multiplets,  $L^{AB+}$ couples to ${T}_{2 \, AB }$ exclusively.
 
 A related argument comes from Sec. 4 where below eq. \rf{act} we explain the next step, namely an integration over $\cC_{1\mu }$. We observe there that `the only term depending on it is $ \cC_{1 \mu} \partial_\nu \tilde  F_2^{\mu\nu} $. This equation is solved if $F_{2 \mu\nu}\approx  \partial_{[\mu } \cB_{2\nu ]}$'. But this would not have worked if we would allow some of the terms in front of the Lagrange multiplier to depend on the 1st multiplet.
 And since the action must depend only on doublets, we cannot include the dependence on just $\cB_{1\mu }$ without $\cC_{1\mu }$.
 
 In case of degenerate groups of type E7 when symmetric bilinear duality invariants are available, the whole procedure has to be revisited since the first term in the 1st order action \rf{ActionDef} is not the only one possible. But our claim with regard to $\cN\geq 5$ supergravities relies  on non-degenerate duality groups of type E7 when the first term in the 1st order action \rf{ActionDef} is  the only one possible, which eliminates the arbitrariness in implementing the deformation
proposal.
   
 Thus the solution of the constraint $ {T}_{AB \, 2}{}^{ + \, def }=0$ is
\be \label{sol2}
 G^+_{2 \La}  = (\cN  F_2)_{ \La} ^+  - 2 \la f^{-1 AB}_\La (\partial_{\mu}\partial_{\nu}\bar f^{-1} F_2^-{})_{ AB}
(\partial^{\mu}\bar f^{-1} F_2^- {})_{ CD}{}
(\partial^{\nu}f^{-1} F_2^{+  CD} ) +\cdots 
\end{equation} 
The first term in $G_2$ is linear in $F_2$ but the deformation term is cubic in $F_2$. Higher order in $\la$ will contain higher powers in $F_2$ in the solution for $G_2$
\subsection{A solution of the deformed ``constraint''  for $G_1$}
For the purpose of deriving the 2d order action from the 1st order one, we could have just observed that all equations of motion are duality covariant since the action is manifestly duality invariant. However, it is interesting to derive the ``constraint'' on $\cF_1$ which in the classical case was derived in Sec. 3 and was shown to be the same as the one for $\cF_2$. 

Deformed  equations of motion take into account that the deformed ``constraint''  is complicated and, for example, ${T}_{2 \, AB }{}^{ + \, def }$ depends not only on $F_2^+$ and $G_2^+$ but also on $F_2^-$ and $G_2^-$
\be
{\delta \cL^{def} \over \delta F_2^{+ \Si}} = 0 \quad \Rightarrow \quad   G_{1\Si}^+ -  L^{ + AB } {\delta T^{+def}_{2 AB } \over \delta F_2^{+ \Si}} + L_{ - AB } {\delta T^{* -def AB}_{ 2} \over \delta F_2^{+ \Si}}= 0 \ ,
\label{G1}\ee
\be
{\delta \cL^{def}  \over \delta G_{2 \Si}^{+ }}=0 \quad \Rightarrow \quad   F_{1}^{\Si +} +      L^{ + AB } {\delta T^{+def}_{AB 2} \over \delta G_2^{+ \Si}} - L_{ - AB } {\delta  T^{*-def AB}_{ 2} \over \delta G_2^{+ \Si}}= 0 \ .
\label{2D}\ee
We can add the conjugate equations and eliminate the Lagrange multipliers, find the solution for them in terms of vectors. It take the following form
$ 
  L^{ + AB } = (Y F_1^+ + Z F_1^-)^{AB}
$.
   Here $Y$ and $Z$ depend on scalars and on $F_2$ and $G_2$. Once the solution for the Lagrange multipliers
 is plugged into  an expression defining $G_1$, we find that
$
 G_{1 \Si}^+ = (V F^+_1 + W F^-_1)_{\Si} =  G_{1 \Si}^+ ( F_1, F_2,  f,h)
$
where $V, W$ now depend  
 on $F_2$ and $G_2$. Note that  the ``constraint''  imposed by the equation for Lagrange multipliers on $\cF_2$ leads to a solution for $G_2$ in terms of $F_2$ only, see eq. \rf{sol2}
 \be
\label{linear1}
  {T}_{2\, AB}{}^{ +} - \la {\delta I (T^-_2, \overline T^-_2)\over \delta \overline T^{AB-}_2}=0  \quad \Rightarrow \quad G_{2 \Si} = G_{2 \Si} ( F_2, f, h) \ .
\ee 
 When this equation  is solved, one finds that $G_{2 \Si}$ depends on scalars and $F_2$, it does not depend on $\cF_1$ by construction since it solves the equation $T^{+def}_2=0$ which depends only on $\cF_2$. In general, 
 one finds that due to a quartic deformation $G_2$ is given by an infinite series in powers of $F_2$. The symmetry between $\cF_1$ and $\cF_2$, as opposite to the classical supergravity, is not restored on shell:
 \be
  G_{2 }^{def} = G_{2 }^{def} ( F_2, f, h)\ , \qquad  G_{1 }^{def} =G_{1 }^{def} ( F_1, F_2, f,h) \ ,
\ee  
 and
 \be
{\delta G_2^{def}\over \delta F_1}=0\ , \qquad  {\delta G_1^{def}\over \delta F_2}\neq 0 \ .
 \ee
This is opposite to the classical case where we have found that
\be
  G_{2 } = G_{2 } ( F_2, f, h)\ , \qquad  G_{1 } =G_{1 } ( F_1, f,h)\ ,  \qquad {\delta G_2\over \delta F_1}={\delta G_1\over \delta F_2}= 0 \ .
\ee  
 Quartic in vectors deformation breaks the on shell symmetry between 
$\cF_1$ and $\cF_2$. This already suggests that we may encounter a related problem in deriving a 2d order action in a deformed theory.

\section{From the 1st to the 2d  order  with  deformation: ghosts do not decouple}
We start with the action in \rf{ActionDef}. As in classical case we integrate the Lagrange multiplier and find $G_2$ as a functional of $F_2$ and scalars. We integrate over $\cC_1$ and identify $F_{2 \mu\nu}$ with $ \partial_{[\mu } \cB_{2\nu ]}$. The terms remaining in the action are
\begin{equation}
  - i G_2^+ F_1^+  +h.c.
 \end{equation}
We use the expression for $G_2$ in \rf{sol2} which we have found by resolving the constraint when integrating over the Lagrange multiplier and find
\begin{equation}
  - i F_1^+  \cN  F^+_2  - 2 \la F_1^{+\La} f^{-1 AB}_\La (\partial_{\mu}\partial_{\nu}\bar f^{-1} F_2^-{})_{ AB}
(\partial^{\mu}\bar f^{-1} F_2^- {})_{ CD}{}
(\partial^{\nu}f^{-1} F_2^{+  CD} )  
    +\cdots    +h.c.
\label{tr2} \end{equation}
We replace as before $ F_1= F + \mathbb{F}$\,, $F_2= F - \mathbb{F}=  F_2$, and find
\be
 - i F^+  \cN  F^+ + i \mathbb{F}^+  \cN  \mathbb{F}^+ 
 - 2 \la (F + \mathbb{F})^+  f^{-1}\partial_{\mu}\partial_{\nu} \bar f^{-1}(F - \mathbb{F})^-
\partial^{\mu} \bar f^{-1}(F - \mathbb{F})^- 
\partial^{\nu} f^{-1}(F - \mathbb{F})^{+  } +h.c.
\ee
where we have skipped other terms $\dots$ and indices for simplicity.
The kinetic terms for normal vectors $F$ and for ghosts vectors $\mathbb{F}$ are as before in a classical case, the normal have a correct sign, the ghosts have a wrong sign. But now, when the deformation terms  are present and $\la\neq 0$, we find interaction terms between normal vectors $F$ and  ghosts vectors $\mathbb{F}$. 

Besides the single term shown here, there are many other terms at the level $\la$, as well as an infinite series of terms with higher and higher powers of $\la$ and increasing powers of vectors. For example, at the level $\la$ one finds in \rf{tomas} terms of a different $S U(8)$ structure. We can present eq. \rf{tomas} in the form
\begin{equation}
\begin{array}{rcl}
& & 
2\mathrm{Tr}_{SU(8)} 
\left(
\partial_{\mu}\partial_{\nu} T^{-}{}_{\alpha\beta}
 T^{*+}{}_{\dot{\alpha}\dot{\beta}}
\right)
\,
\mathrm{Tr}_{SU(8)} 
\left(
\partial^{\mu}T^{-\, \alpha\beta}
\partial^{\nu}T^{*+\, \dot{\alpha}\dot{\beta}}
\right)
\\
& & \\
& & 
+
\mathrm{Tr}_{SU(8)} 
\left(
\partial_{\mu}\partial_{\nu}T^{-}{}_{\alpha\beta}
\partial^{\mu}T^{*+}{}_{\dot{\alpha}\dot{\beta}}
T^{-\, \alpha\beta}
\partial^{\nu}T^{*+\, \dot{\alpha}\dot{\beta}}
\right)
+\mathrm{c.c.}
\end{array}
\label{tomas1}\end{equation}
The term linear in $\la$ in eq. \rf{tr2} originate from the first term in eq. \rf{tomas1}, of the symbolic form $(\mathrm{Tr}_{SU(8)} TT)^2$. The terms linear in $\la$ in $\dots$ in eq. \rf{tr2} include the ones which originate from  the second term in eq. \rf{tomas1}, of the symbolic form $(\mathrm{Tr}_{SU(8)} TTTT)$. We have not given the details on this term, but they clearly can only add to an interaction terms between normal vectors $F$ and  ghosts vectors $\mathbb{F}$, they cannot cancel the terms shown in eq. \rf{tr2}, since they have a different $SU(8)$ nature. For $\cN<8$ where the indices $A, B$ take values $1, \dots, \cN<8$ the same property is inherited via truncation, one still finds different $SU(\cN)$ structure terms which cannot cancel each other.

Moreover, 
we have presented an example of a possible source of deformation in eq. \rf{tomas}, associated with the  3-loop candidate counterterms. Many other sources of the deformation are also possible, with additional derivatives, associated with higher loop candidate counterterms or with a different contraction of indices. But as we see in the simplest example the reason for ghosts non-decoupling is that the deformation is quartic in vectors. The number of derivatives and the choice of contraction of indices do not matter.

For example, at the 4-loop level the candidate counterterm suggesting a source of deformation at the 4-vector loop level has the same form as the one in \rf{tomas} but in addition has two extra space-time derivatives, acting on the 4 vectors. The exact form of this 4-loop  candidate counterterm defining the source of deformation can be established for example using the amplitude computations as it was done following \cite{Freedman:2011uc}, where the details  for the 3-loop case are presented.

All these terms have normal vectors coupled to ghosts vectors.
Thus, the ghosts do not decouple, and therefore the 2d order action following from the 1st order one with manifest E7 type symmetry is inconsistent if local candidate counterterms deform the action and the constraint.

One may ask the question: Is it possible that in the presence of $\lambda$ the ghost field is not anymore $F_1-F_2$ but it picks up some $\lambda$-dependent term\footnote{We are grateful to R. Roiban for this question. } and this more complicated expression for the deformed ghost field $ \mathbb{F}^{def}$ decouples? 

Note that we have an action where $F_{1 \mu\nu}$ from the beginning was $ \partial_{[\mu } \cB_{1\nu ]}$, and after integrating the 
action over $\cC_{\mu1}$ using $ \cC_{\mu1} \partial_\nu \tilde  F_2^{\mu\nu} $ we found that $F_{2 \mu\nu}= \partial_{[\mu } \cB_{2\nu ]}$. Thus we look for a deformation of the ghost vector field like
\be
 \mathbb{F}_{ \mu\nu}^{def} = F_{1 \mu\nu}-  F_{2 \mu\nu} =\partial_{[\mu } \cB_{1\nu ]}-  \partial_{[\mu } \cB_{2\nu ]} +\la \, \partial_{[\mu } \cB_{3\nu ]}\ ,
 \ee
so that it can be implemented via the change of variables in the functional integral, where we integrate over $\cB_{1\nu }$ and $\cB_{2\nu }$.
\be
\cB_{1\nu } - \cB_{2\nu } \Rightarrow \cB_{1\nu } - \cB_{2\nu } + \la \cB_{3\nu } \ .
\ee
To decouple the vector ghost we need to find out if from the four vector coupling term in the action depending on $ \mathbb{F}$ we can extract  the expression in the form of $\partial_{[\mu } \cB_{3\nu ]}$, so that in the first approximation in $\la$ we can absorb the four-vector term into a deformation of the kinetic term. The four vector can be given in the form $ \mathbb{F}_{ \mu\nu}^+ X^{\mu\nu +} +cc$. The  expression for $X^{\mu\nu}$ is cubic in $F$ and $ \mathbb{F}$, and function of scalars. To be able to absorb this terms into a redefinition of the kinetic term we have to show that $X^{\mu\nu +} = \cN Y^{\mu\nu +}$ where $Y_{\mu\nu}= \partial_{[\mu } \cB_{3\nu ]}  $. However, $Y_{\mu\nu}$ is a complicated expression cubic in vectors and it is not of the form $\partial_{[\mu } \cB_{3\nu ]}$ off shell. Therefore the decoupling of  vector ghosts is not possible, once $\la\neq 0$.

So  far we have shown that the 4-vector deformation corresponding to any particular loop candidate counterterm, starting with $L=3$ will break  a non-degenerate type E7 symmetry of the theory. However, without supersymmetry, this observation is not very useful: one can see from eq. \rf{con} that any candidate for UV divergence which is independent of vectors, escapes our argument that it should not show up when computing UV divergences. For example, the 82 diagrams in  \cite{Bern:2014sna} represent 4-point amplitudes: these describe 4-graviton amplitude, 4-vector amplitude etc.
Only when supersymmetry is unbroken, our analysis of the 4-vector amplitude becomes a relevant analysis of the superamplitude, and we can conclude that E7 type supersymmetry and supersymmetry protect $\cN\geq 5$ supergravity from UV loop divergences.

\section{Discussion}

Our main result here is the derivation of the 1st order action with manifest duality symmetry of the E7 type for classical $\cN\geq 5$ supergravities given in eq. \rf{Action}. 
The action is based on a universal symplectic approach used recently in  \cite{Kallosh:2018mlw} 
to study the deformation of supergravities due to candidate counterterms in perturbative theory. It was 
developed earlier in the context of  supersymmetric black hole attractors in \cite{Andrianopoli:1996ve}. The defining feature of this action is that the first term in it, a bilinear invariant
\be
\langle  \cF_1\mid \tilde  \cF_2\rangle  \equiv
\tilde  \cF_2^{\Lambda}\cF_{ 1\Lambda} 
-\tilde  \cF_{ 2\Lambda}\cF_1^{\Lambda}\, ,
\ee
vanishes for a single duality doublet; only an antisymmetric in two doublets invariant is possible for non-degenerate E7 type groups in $\cN\geq 5$ supergravities.

 Our 1st order action \rf{Action} is valid for the most interesting maximal supersymmetry case of $\cN=8$, as well as for $\cN=5$, where there is an information about 4 loop UV finiteness in d=4,  which was not explained until now. In $\cN=8$ case we have reproduced the main result of Cremmer and Julia  \cite{Cremmer:1979up}:  when deriving 2d order action from the manifestly \E\, invariant 1st order action, one encounters ghosts, but they decouple. This renders the classical theory without additional local terms in the action, associated with UV divergences, ghosts-free and preserving E7 type symmetry for all $\cN\geq 5$ supergravities.

We also found here that when the theory is deformed via a local 4-vector candidate counterterm,  the ghosts of the 2d order action, following from the E7 invariant 1st order action,  do not decouple.   But other terms, like any vector independent 4-point deformation due to a candidate UV divergence, are not 
forbidden  by the E7 type argument  here. Thus our duality symmetry argument forbids only one of the possible various 4-point candidate counterterms. 

Other 4-point terms, there are about 50 of them, are supersymmetric partners
 of the 4-vector candidate UV divergences. If supersymmetry is unbroken,  all of them are also forbidden since it requires all 4-point terms in a given candidate UV divergence to come up in perturbative loop computations with the same factor as the one in the 4-vector case.\footnote{An  investigation in \cite{Gunaydin:2018kdz}  shows an inconsistency between the deformation proposal  \cite{Bossard:2011ij} and linearized supersymmetry. It  suggests an independent explanation of 4 loop $\cN=5$  UV finiteness.}

It is  important to stress that our  E7 type duality-supersymmetry argument for the absence of UV divergences is valid for $\cN\geq 5$ perturbative supergravity at any loop order. Its validity depends only on  the validity of our  assumption that both duality and supersymmetry have no anomalies when loop computations are performed.

In particular, our analysis suggests an  explanation of  the mysterious cancellation of the 82 diagrams in $\cN=5$, 4 loop theory in four dimensions, discovered in \cite{Bern:2014sna}.  As far as we know, no other explanation of this cancelation was proposed during the last four years since the result in \cite{Bern:2014sna} was published.\footnote{Note that  duality symmetry requires a  vanishing soft limit on amplitudes with scalars, but this  does not explain UV finiteness of $\cN=5$  at 4 loops  \cite{Freedman:2018mrv}, same as for $\cN=8$ at 7 loops. However, duality symmetry in the vector sector, as we argued in this paper, does explain it. }  Our analysis here implies that if at the level of 4-loop  in d=4, $\cN=5$ theory has no duality and no supersymmetry anomalies,  it has to be UV finite at this level.

In conclusion, our analysis of duality symmetry,  in absence of duality and supersymmetry anomalies,  when the predictions of these symmetries are respected in quantum corrections,
 suggests that $\cN\geq 5$ supergravities may be perturbatively UV finite.

\

 \noindent{\bf {Acknowledgments:}} I am grateful to S.~Ferrara, D.~Freedman, M.~G\"unaydin, A.~Linde, H.~Nicolai,  A.~Tseytlin  and  Y.~Yamada   for stimulating discussions and collaboration on related work, and I thank R.~Roiban,  A.~Van~Proeyen and B.~de~Wit for the valuable comments on the draft.  I am grateful to the Editor and to the Referee for suggestions how to clarify and improve the paper.
 This work is supported by SITP and by the US National Science Foundation grant PHY-1720397.


\appendix 

\section{Supergravity and groups of type E7  }

We present the duality groups  $\cG$ of  the corresponding ${\cG\over \cH}$ symmetric spaces   and their symplectic representations  $\mathbf{R}$
of $\cN\geq 3$ supergravities in Table 1 and the ones in $\cN=2$ in Table 2.
  All extended $\cN>1$ supergravities  which are described by symmetric coset spaces ${\cG\over \cH} $  have non-degenerate duality groups $\cG$ of type E7, with exception of degenerate 
$U(p,n)$ models. In the past this fact was used either in the context of black hole attractors or, more recently,  in the cosmological context in \cite{Ferrara:2011dz,Ferrara:2012qp},  where also many earlier references are discussed, in particular \cite{Borsten:2011nq}.

Our interest here is different both from the black hole as well as  cosmology studies and, therefore, we have to explain  the mathematical aspect  of groups of type E7 relevant in our context of a manifestly E7 invariant actions. 
The appearance of ghosts in an action with a  manifest E7 type symmetry is only due to the absence of a symmetric quadratic duality invariant, as explained in sec. 4.1.

The concept of Lie groups of type E7 was introduced in  1967 by Brown  \cite{Brown}, and then later developed in both mathematical as well as physical literature, e.g. in \cite{Garibaldi,Borsten:2011nq,Ferrara:2011dz,Ferrara:2012qp}.
The E7 type groups are defined by an irreducible linear representation with only two primitive invariants: a symplectic form $\langle  \mathcal{A}\mid \mathcal{B}\rangle \equiv
  \mathcal{B}^{\Lambda}\mathcal{A}_{\Lambda} 
-\mathcal{B}_{\Lambda}\mathcal{A}^{\Lambda} $ and a symmetric quartic invariant: 
\begin{equation}
\mathbf{q}\left( \mathcal{Q}\right) \equiv \varsigma \mathbb{K}_{MNPQ}%
\mathcal{Q}^{M}\mathcal{Q}^{N}\mathcal{Q}^{P}\mathcal{Q}^{Q},  \label{I4}
\end{equation}%
A famous example of \textit{quartic} invariant in $\cG=E_{7(7)}$ is the \textit{%
Cartan-Cremmer-Julia} invariant \cite{Cremmer:1979up},  constructed out of
the fundamental representation $\mathbf{56}$.

A  more rigorous  definition of groups of type E7 involves 3 features. Here we literally present a definition as given by Brown in 1969 in 
\cite{Brown} as well as almost 50 years later in \cite{Borsten:2018djw}.

Groups of type $E_7$ can be characterized by Freudenthal triple systems (FTS). A FTS may be axiomatically defined  as a finite
dimensional vector space $\mathfrak{F}$ over a field $\mathfrak{J}$ (not of
characteristic 2 or 3), such that:
\begin{enumerate}
\item  $\mathfrak{F}$ possesses a non-degenerate antisymmetric bilinear form $\{x, y\}.$
\item $\mathfrak{F}$ possesses a symmetric four-linear form $q(x,y,z,w)$ which is not identically zero.
\item If the ternary product $T(x,y,z)$ is defined on $\mathfrak{F}$ by $\{T(x,y,z), w\}=q(x, y, z, w)$, then
\be
3\{T(x, x, y), T(y,y,y)\}=\{x, y\}q(x, y, y, y).
\ee
\end{enumerate}
We refer to more definitions,  details and earlier references to the most recent work on groups of type E7 in the context of the Freudenthal duality and black holes in  \cite{Borsten:2018djw}.

For our purpose of analysis of these groups in supergravity an important progress in understanding groups of type E7  is due to Garibaldi \cite{Garibaldi}. He studied 
groups of type E7 over arbitrary fields, with  characteristic $\neq 2, 3$, including real-closed field.
 He has noticed that FTS comes in two flavors, degenerate and non-degenerate.  He stressed that the FTS is non-degenerate precisely when the quartic form is irreducible, and degenerate otherwise. In Table 1 we present simple, \textit{non-degenerate} duality groups in supergravity related to FTS.

\begin{table}[t]
\begin{center}
\begin{tabular}{|c||c|c|c|}
\hline
$%
\begin{array}{c}
J_{3}%
\end{array}%
$ & $%
\begin{array}{c}
G_{4} \\
\end{array}%
$ & $%
\begin{array}{c}
\mathbf{R} \\
\end{array}%
$ & $%
\begin{array}{c}
\mathcal{N} \\
\end{array}%
$ \\ \hline\hline
$%
\begin{array}{c}
J_{3}^{\mathbb{O}} \\
\end{array}%
$ & $E_{7\left( -25\right) }~$ & $\mathbf{56}$ & $2~$ \\ \hline
$%
\begin{array}{c}
J_{3}^{\mathbb{O}_{s}} \\
\end{array}%
$ & $E_{7\left( 7\right) }$ & $\mathbf{56}$ & $8$ \\ \hline
$%
\begin{array}{c}
J_{3}^{\mathbb{H}} \\
\end{array}%
$ & $SO^{\ast }\left( 12\right) $ & $\mathbf{32}$ & $2,~6$ \\ \hline
$%
\begin{array}{c}
J_{3}^{\mathbb{C}} \\
\end{array}%
$ & $SU\left( 3,3\right) $ & $\mathbf{20}$ & $2~$ \\ \hline
$%
\begin{array}{c}
M_{1,2}\left( \mathbb{O}\right) \\
\end{array}%
$ & $SU\left( 1,5\right) $ & $\mathbf{20}$ & $5$ \\ \hline
$%
\begin{array}{c}
J_{3}^{\mathbb{R}} \\
\end{array}%
$ & $Sp\left( 6,\mathbb{R}\right) $ & $\mathbf{14}^{\prime }$ & $2$ \\ \hline
$%
\begin{array}{c}
\mathbb{R} \\
(T^{3}\text{~model})~~%
\end{array}%
$ & $SL\left( 2,\mathbb{R}\right) $ & $\mathbf{4}$ & $2$ \\ \hline
\end{tabular}%
\end{center}
\caption{Simple, \textit{non-degenerate} duality groups $\cG= G_4$ in $d=4$ related to
FTS $\mathfrak{M}\left( J_{3}\right) $ on simple rank-%
$3$ Jordan algebras $J_{3}$ with   symplectic irreducible representations. $\mathbf{R}$.  Here $\mathbb{O}$, $\mathbb{H}$, $\mathbb{C}$ and $%
\mathbb{R}$ respectively denote the four division algebras of octonions,
quaternions, complex and real numbers, and $\mathbb{O}_{s}$, $\mathbb{H}_{s}$%
, $\mathbb{C}_{s}$ are the corresponding split forms. The $G_{4}$
related to split forms $\mathbb{O}_{s}$, $\mathbb{H}_{s}$, $\mathbb{C}_{s}$
is the \textit{maximally non-compact} (\textit{split}) real form of the
corresponding compact Lie group.  $J_{3}^{\mathbb{H}}$ is related
to both $8$ and $24$ supersymmetries, because the corresponding supergravity
theories are \textit{\textquotedblleft twins"},  they share the 
same bosonic sector. }
\end{table}

Garibaldi \cite{Garibaldi}  also studied the FTS  with  the central simple algebra component split. He has presented the cases with a degenerate FTS such that 
 \be
q(x, x, x, x) := 12  \det (x)^2
 \ee
The case when there is  a symmetric quadratic invariant form,  constructible from the other  invariants is therefore known as a
``degenerations'' of the duality groups of type E7.  These include cases when the corresponding quartic invariant polynomial built from the symplectic irreducible representation  	``degenerates" into a perfect square. In  \cite{Ferrara:2011dz} these cases were called `not E7 type',  for simplicity, whereas in \cite{Ferrara:2012qp} we have followed the terminology in \cite{Garibaldi} and called them cases with ``degenerations'' of the duality groups of type E7. 
\begin{table}[t]
\begin{center}
\begin{tabular}{|c||c|c|}
\hline
$\cN$& $
\begin{array}{c}
$G$ \\
\end{array}
$ & $
\begin{array}{c}
  \mathbf{ R}
\end{array}
$ \\ \hline\hline
$
\begin{array}{c}
3 \\
\end{array}
$ & $U(3,n)$ & $ \mathbf{(3+n)}$     \\ \hline
$
\begin{array}{c}
4 \\
\end{array}
$ & $SL(2, \mathbb{R})\otimes {SO(6,n)}$ & $\mathbf{(2, 6+n)}$   \\ \hline
$
\begin{array}{c}
5 \\
\end{array}
$ & $SU(1,5)$ & $ \mathbf{ 20}$   \\ \hline
$
\begin{array}{c}
6 \\
\end{array}
$ & $SO^{\ast }(12)$ & $\mathbf{ 32}$
 \\ \hline
$
\begin{array}{c}
8 \\
\end{array}
$ & $E_{7\left( 7\right) }$ & $\mathbf{ 56}$   \\ \hline
\end{tabular}
\end{center}
\caption{ $\cN\geqslant 3$ supergravity sequence of groups $\cG$ of  the corresponding ${\cG\over \cH}$ symmetric spaces, and their symplectic representations  $\mathbf{R}$. Note that in the case of $\cN=4$  duality group is non-degenerate, yet it is anomalous.}
\end{table}

\begin{table}[t]
\begin{center}
\begin{tabular}{|c||c|}
\hline
$
\begin{array}{c}
$G$\\
\end{array}
$ & $\mathbb{\mathbb{}}
\begin{array}{c}
  \mathbf{ R}\\
\end{array}
$ \\ \hline\hline
$
\begin{array}{c}
{U(1,n)}
\end{array}
$ & $
\begin{array}{c}
\mathbf{(1+n)_c}\\
\end{array}
$ \\ \hline
$
\begin{array}{c}
{SL(2, \mathbb{R})}\otimes SO(2,n)
\end{array}
$ & $
\begin{array}{c}
\mathbf{(2, 2+n)}
\end{array}
$ \\ \hline
$
\begin{array}{c}
SL(2, \mathbb{R})
\end{array}
$ & $
\begin{array}{c}
\mathbf{4}
\end{array}
$ \\ \hline
$
\begin{array}{c}
Sp(6,\mathbb{R})~
\end{array}
$ & $
\begin{array}{c}
\mathbf{14}' \\
\end{array}
$ \\ \hline
$
\begin{array}{c}
SU(3,3)\end{array}
$ & $
\begin{array}{c}
\mathbf{20}
\end{array}
$ \\ \hline
$
\begin{array}{c}
SO^{\ast }(12)~
\end{array}
$ & $
\begin{array}{c}
\mathbf{32}
\end{array}
$ \\ \hline
$
\begin{array}{c}
E_{7\left( -25\right) }
\end{array}
$ & $
\begin{array}{c}
\mathbf{56}
\end{array}
$ \\ \hline
\end{tabular}
\end{center}
\caption{$\cN=2$ choices of groups $\cG$ of the  ${\cG\over \cH}$ symmetric spaces and their symplectic representations  $\mathbf{R}$. The last four lines refer to ``magic $\cN=2$ supergravities''.}
\end{table}

The meaning of ``degenerate" groups of type E7 is that, in notation of  \cite{Ferrara:2012qp}, the quartic invariants have the property that 
\begin{equation}
\mathbb{K}_{MNPQ}=\frac{\zeta ^{2}}{3}\mathbb{S}_{M(N}\mathbb{S}_{PQ)}.
\label{K-deg}
\end{equation}%
where $\mathbb{S}_{(PQ)}$ is a  rank-2 symmetric invariant symplectic tensor. 
By
introducing
\begin{equation}
\mathcal{I}_{2}\left( x,y\right) \equiv \zeta \mathbb{S}_{MN}\mathcal{Q}%
_{x}^{M}\mathcal{Q}_{y}^{N},
\end{equation}%
one can  check, see \cite{Ferrara:2012qp} for  details,  that the  corresponding quartic invariant 	``degenerates" into a perfect square:
\begin{gather}
-{1\over 6} \mathbf{q}\left( x,y,z,w\right) \equiv \mathbb{K}_{MNPQ}\mathcal{Q}_{x}^{M}%
\mathcal{Q}_{y}^{N}\mathcal{Q}_{z}^{P}\mathcal{Q}_{w}^{Q}\Big |_{x=y=z=w}  \Rightarrow  \,  [ \mathcal{I}_{2}\left(
x,x\right) ]^{2} \ .  \label{q-x-x-x-x}
\end{gather}
It has been verified in \cite{Ferrara:2011dz,Ferrara:2012qp}, that the duality groups 
$\cG = U(r,s)$ which describe some of $\cN=3,2$ supergravities belong to ``degenerate" groups of type E7, or  are `not  groups  E7', since they have a symmetric quadratic invariant. In such case our analysis above  does not apply.

We show in Table 2 the duality groups of  $\cN\geq 3$ supergravities interacting with $n$ vector multiplets, when available. Only the ones for $\cN=3$ are degenerate, or not of E7 type, the rest, 
with $\cN\geq4$ is not degenerate. It means that the 4-vector deformation would break $\cG$-symmetry for $\cN\geq 4$, which should not happen if 
  the symmetry is not anomalous and controls quantum corrections.

We also provide Table 3 where choices of duality group are given for  $\cN= 2$ supergravities interacting with vector multiplets. Only the models in first line with $\cG= U(1,n)$ would be not be protected by dualities since  ghosts are not shown to be present in $\cG = U(r,s)$ case. 
There is an interesting situation with the so-called `magical' N=2 supergravities \cite{Gunaydin:1983rk,Gunaydin:1983bi}, described in Table 2 in the last 4 lines.
The U-duality groups of these four magical supergravity theories in $d=4$ are all groups of type E7, namely  $E_{7(-25)} , SO^*(12) , SU(3,3)$ and $Sp(6,\mathbb{R})$. Their U-duality groups are simple and the vector field strengths and their magnetic duals  form a single irreducible symplectic representation. This is a property  they share with $\cN\geq 5$ supergravity theories.
However unlike $\cN \geq 5$  supergravities generic $\cN=2$ Maxwell-Einstein supergravity theories with homogeneous scalar manifolds have one loop divergences. These divergences correspond to two independent linearized  counterterms and the divergences associated with one of these counterterms are absent {\it only }  for the magical supergravity theories, as shown in \cite{Ben-Shahar:2018uie}. The first UV divergence in \cite{Ben-Shahar:2018uie} corresponds to the term $(T_{\mu\nu}^{ \rm mat})^2 +\cdots $. It is duality invariant since the energy momentum tensor is duality invariant. The fact that in magical supergravities the second type of UV divergence vanishes might be a consequence of a non-degenerate E7 type duality.

However, same as in the case of models in Table 2,  an analysis of duality symmetry groups with regard to UV properties is not complete: we have to look at the situation with anomalies to  decide about UV properties of $\cN\leq 4$ supergravities.

\section{Anomalies in supergravity}
It was known early  on from \cite{Marcus:1985yy}, that chiral $U(1)$ anomalies are always present in $\cN<5$. These $U(1)$ anomalies belong to  some subgroup of a duality group $\cG$. Therefore from the beginning one would have expected that dualities may not be controlling quantum corrections in $\cN<5$ supergravities. But now we know much more. In particular, it was observed in \cite{Meissner:2016onk} that the so-called `conformal anomalies' defined by  
\be
{\cal A}_{\rm conf}=T^\mu_{\; \; \mu}=\frac{1}{180(4\pi)^2}\left(c_{s}\cC^2+a_s \GB \right) ,
\label{anom}
\ee
where 
\bea\label{CGB}
\cC^2&\equiv&\cC_{\mu\nu\rho\si}\cC^{\mu\nu\rho\si}=R_{\mu\nu\rho\si}
R^{\mu\nu\rho\si}-2R_{\mu\nu}R^{\mu\nu}+\frac13 R^2 \ ,\nn\\
\GB &=& R^* R^*= R_{\mu\nu\rho\si}
R^{\mu\nu\rho\si}-4R_{\mu\nu}R^{\mu\nu}+R^2 \ ,
\eea
are given by Table 4. The coefficients $c_s$ and $a_s$ depend on the spin $s$ of the fields that
couple to gravity.
\renewcommand{\arraystretch}{1.3}
\begin{table}[t]
\begin{center}
\begin{tabular}{|c||c|c||}
\hline
& $c_s$ & $a_s$\\ \hline \hline
\  $0$($0^*$) &\ $\frac32$$(\frac32)\ $ &\ $-\frac12$($\frac{179}2$)  \\[3pt] \hline
$\frac12$ & $\frac92$  & $-\frac{11}{4}$  \\[1pt] \hline
$1$ & $18$  & $-31$   \\[1pt] \hline
$\frac32$ & $-\frac{411}{2}$   & $\frac{589}{4}$    \\[1pt] \hline
$2$ & $783$   & $-571$     \\[1pt] \hline
\end{tabular}
\end{center}
\caption{Coefficients of the conformal anomaly in \cite{Meissner:2016onk}.  The entry labeled $0^*$ give the result
for two-form field; it gives the same contribution 
to $c_0$ as the scalars, but its contribution to  the $a_0$ coefficient is different.}
 \end{table}

Here $\GB$ is the Gauss-Bonnet density and 
$\cC^2$ is the square of the Weyl tensor.  Only gravitino have negative contribution to $c_s$, see Table 3,  and therefore  
the cancellation of $c_s$ takes place for $\cN=5,6,8$ but in all cases below $\cN=5$ there is no cancellation, \cite{Meissner:2016onk}.  For pure $\cN$ supergravities one finds
\bea\label{5}
{\cal A}_{\rm conf}^{\cN=4}&= & c_2 + 4 c_{\frac32} + \, 6 c_1 \, + \, \, 4 c_{\frac12} \, + \, \, 2 c_0 = 90\; ,\\
{\cal A}_{\rm conf}^{\cN=5}&=&  c_2 + 5 c_{\frac32} + 10 c_1 + 11 c_{\frac12} + 10 c_0 = 0\; ,\\
\label{6}
{\cal A}_{\rm conf}^{\cN=6}&=& c_2 + 6 c_{\frac32} + 16 c_1 + 26 c_{\frac12} + 30 c_0 = 0 \; ,\\
\label{8}
{\cal A}_{\rm conf}^{\cN=8}&=& c_2 + 8 c_{\frac32} + 28 c_1 + 56 c_{\frac12} + 70 c_0 = 0 \;.
\eea
Adding matter multiplets will add  the contribution from $c_0, c_{1/2}, c_{1}$ to $\cC^2$ anomaly, these are all positive, the value of this anomaly in presence of  matter can only grow in value, not cancel the one from pure supergravity.

It has been discussed in \cite{Kallosh:2016xnm} that the  cancellation of conformal anomalies in $d=4$ for  $C_{\alpha \beta \gamma\delta} C^{\alpha \beta \gamma\delta} +\bar C_{\dot \alpha \dot \beta \dot \gamma \dot \delta} \bar C^{\dot \alpha \dot \beta \dot \gamma \dot \delta}$  and  
 of chiral anomalies for the $C_{\alpha \beta \gamma\delta} C^{\alpha \beta \gamma\delta} -\bar C_{\dot \alpha \dot \beta \dot \gamma \dot \delta} \bar C^{\dot \alpha \dot \beta \dot \gamma \dot \delta} $ is a property of    $\cN$-extended supergravities with $\cN \geqslant 5$.
It can be explained  using supersymmetry and dimension of linearized chiral superfields. This is in  contrast with  $\cN < 5$ models, where both types of anomalies are known to be present.

The  $\cN=4$ supergravity, pure and matter interacting, is known to have both chiral and well as conformal anomalies, moreover, the 1-loop $U(1)$ anomalous superamplitudes were found in \cite{Carrasco:2013ypa}. This anomalous $U(1)$ is a subgroup of the duality group $SL(2, \mathbb{R})\otimes {SO(6,n)}$, which is a non-degenerate type E7 group, but still anomalous.
In $\cN=4$ at 4-loop the UV divergent amplitudes \cite{Bern:2013uka} have a structure  closely related to the structure of the 1-loop anomalous amplitudes. It was suggested in  \cite{Bern:2017rjw} how to eliminate the 1-loop anomaly. It remains unclear how this procedure will affect the perturbative UV behavior of $\cN=4$ supergravity. 

Meanwhile, it was established in  \cite{Freedman:2017zgq} that in $\cN\geq 5$ the one-loop anomalies in superamplitudes are absent, which is in agreement with the early analysis in \cite{Marcus:1985yy} about chiral anomalies. Therefore the fact that  $\cN=5$ at $L=4$ is not UV divergent also agrees with the facts that 1) duality symmetry of $\cN=5$ is a non-degenerate type E7 group, which according to our observations in this paper, protects the theory from deformations/UV divergences. This protection is reliable since there are no known anomalies.

Here we remind that so far the issue of supergravity anomalies was studied only in the context of the 2d order theory, in the 1st order formalism advocated here, it might have to be developed additionally.

\section{ $\cN\leq 4$ supergravities}

Not all extended $\cN>1$ supergravities are  expected to be UV finite. To understand better the difference between low $\cN$ theories with  $\cN \geq 5$ models, which we argue here are perturbatively  UV finite in absence of anomalies, we would like to describe here also what is known about models  with  $\cN\leq 4$, in addition to what was explained in Appendix A.

First of all, whenever there is a pure supergravity without a matter multiplet, the UV behavior is known to be better than in cases with matter multiplets. In $\cN\geq 5$ there are only gravitational multiplets, no matter. In models with $\cN\leq 4$ there are matter multiplets. When there is matter, already at the 1st loop order the UV divergences are known to be present, for example 
of the form $(T_{\mu\nu}^{ \rm mat})^2 +\cdots $ where $T_{\mu\nu}^{ \rm mat}$ is the energy-momentum tensor of matter. These   vanish in pure supergravities with $T_{\mu\nu}^{ \rm mat}=0$.
 Secondly, it is known that the presence of matter multiplets never improves the status of anomalies (but naive field dualization can) comparative to the one in pure supergravity, as we explained in the case of conformal anomalies  above. Therefore our main interest from the point of view of UV divergences is with pure supergravities. However, we have also discussed above the theories with matter multiplets, their corresponding duality groups as well as anomalies.
Note that in $\cN=1,2$ the scalar field manifold can be inhomogeneous, therefore these models are not associated with coset spaces ${\cG\over \cH}$ 
which we study here.

One has to  find out if duality symmetry has anomalies. For example, in $\cN=3$ there are anomalies, therefore the fact that duality group $SU(3,n)$ is a degenerate type E7 does not matter much, since anomalies make the symmetry not reliable anyway. In $\cN=2$ there are anomalies, therefore despite we see from Table 3 that all  models but the ones with $SU(1,n)$  have non-degenerate duality groups of the type E7,  this symmetry is not reliable, all of them are not UV finite. We refer to a more detailed review of $\cN=2$ supergravities interacting with matter to an Appendix in  \cite{Gunaydin:2018kdz}. Thus, due to anomalies all $\cN < 4$ supergravities are not expected to be perturbatively UV finite.  The status of $\cN=4$ is still to be established.

\begin{figure}[h!]
\begin{center}
\includegraphics[scale=0.6]{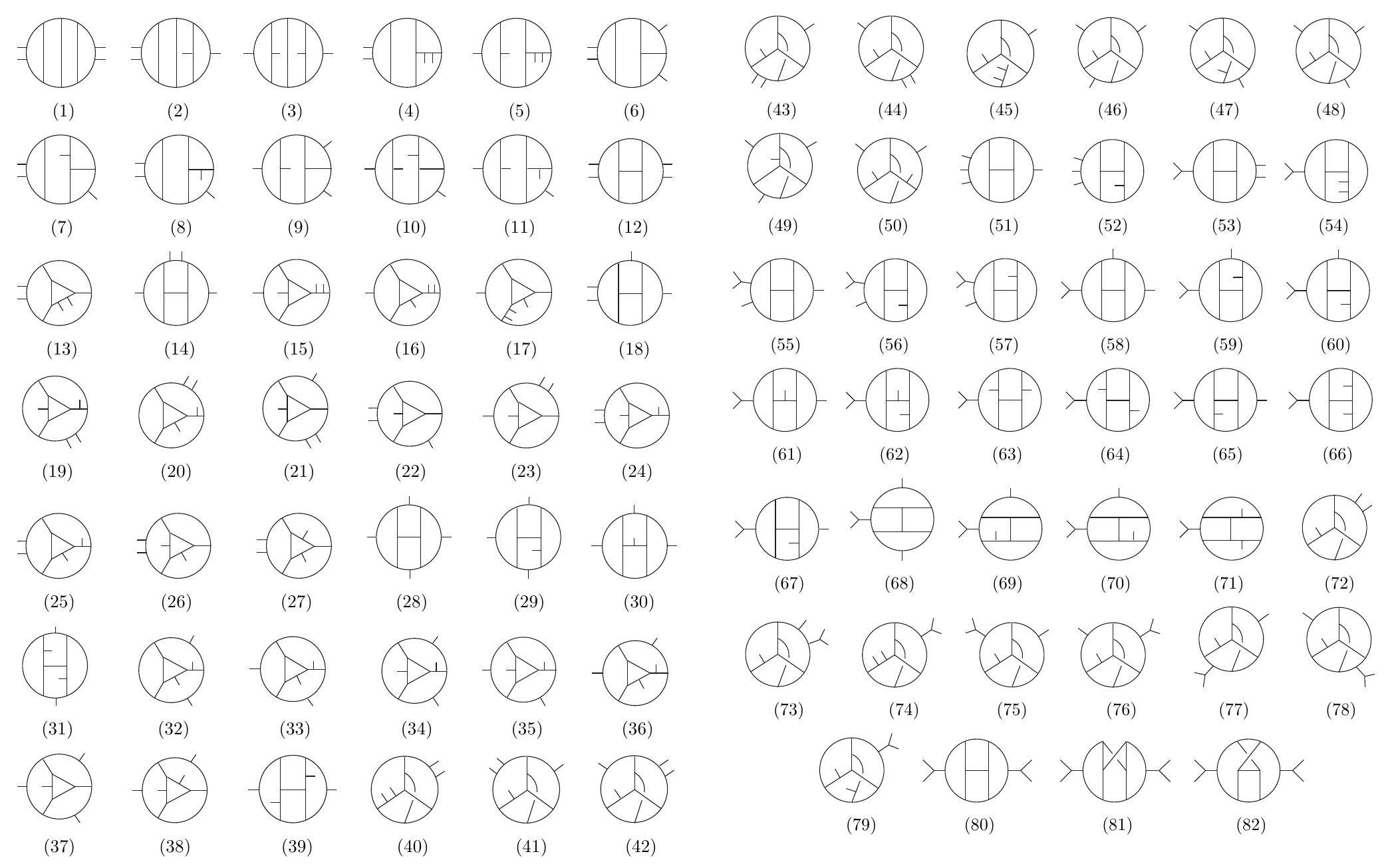}
\end{center}
\caption{\footnotesize 82 diagrams in $\cN=5$, 4 loops.  The individual diagrams are UV divergent in d=4, but  the sum  of all diagrams has no UV divergences   \cite{Bern:2014sna}. }
\label{FO}
\end{figure}

\section{Cancellation of  UV divergences  in 82 diagrams at $L=4$ in $\cN=5$ }
In Fig. 1 we show the set of diagrams from  \cite{Bern:2014sna} where the UV divergences cancel. We suggest that the reason why the UV infinities in this set of diagrams cancel is explained above using the action with a manifest non-degenerate E7 symmetry.

\

\section{ A status of the  BN deformation proposal  in supergravity}

In standard 2d order supergravity  the vector part of the action does not have duality  symmetry. This symmetry  rotates vector field equations into Bianchi identities.
In the second order formalism these are treated in an asymmetric way: the action depends  on $n_v= (28, 16,10)$ vector potentials $\cB_\mu$ via $F= d\cB$.
The Bianchi identity $dF=0$ for $F= d\cB$  are valid off-shell,  whereas equations of motion $dG=0$ with $\tilde G ={1\over 2}  {\delta L\over \delta F} $ are only valid on shell. Therefore $G\approx d\cC$ only in virtue of field equations. The  dual vector $\cC_\mu$ is not present in the action, $G$ is the function of $F$ and scalars and fermion fields,  and the analysis of duality symmetry in the second order formalism relies on the fact that $\delta S= \int GB\tilde G$.

The proof of duality current conservation in \cite{Gaillard:1981rj,Kallosh:2008ic} is somewhat tricky since the 2d order action is not a `bona fide' duality invariant action  and therefore   the proof of the relevant Noether-Gaillard-Zumino current conservation is not transparent. It requires that $\int Gb\tilde G$ vanishes on shell, where $b$ is the infinitesimal part of $B$ in eq. \rf{dualrot}. On the other hand, there is also   a  Noether-Gaillard-Zumino identity \cite{Gaillard:1981rj,Aschieri:2008ns,Kallosh:2011dp} in supergravity:
\be
 \int d^4x\, Gb\tilde{G} =\int d^4x \left[\delta_b \varphi\, \frac{\delta \cL_v}{\delta \varphi}  +h.c.  \right]  \ .
\label{NGZ0}\ee
Here $\cL_v$ is a vector dependent part of the action. The details and examples are presented in Appendix F.
Clearly, in absence of scalars, the right hand side of this identity vanishes, but not when the scalars in ${\cG\over \cH}$ coset space are present. In \cite{Bossard:2011ij},  the action is expanded in a deformation parameter $\la$
\be
S= S_{cl} +\la S^{(1)} + \la^2 S^{(2)} +\cdots
\ee
Here $S^{(1)}$ is the candidate counterterm which is known  \cite{Kallosh:2011dp} to violate the condition  $\int d^4x\, Gb\tilde{G} $ at the $\la^2$ level. The terms of the order $\la^n$ in the proposal \cite{Bossard:2011ij}, indeed, are capable of restoring the condition $\int d^4x\, Gb\tilde{G} =0$ order by order in absence of scalars.

The proof of the consistency of the deformed theory is presented in Appendix of \cite{Bossard:2011ij}.
It is explained there that the dependence on scalars was suppressed  for simplicity. However, looking at eq. \rf{NGZ0} we see now that in absence of scalars the expansion in $\la$ was guaranteed to provide $\int d^4x\, Gb\tilde{G} =0$ order by order in $\la$, but this statement is correct only in absence  of scalars.   Meanwhile, for extended supergravities where scalars are present in the coset space ${\cG\over \cH}$ and duality groups are of the type E7, the right hand side of eq. \rf{NGZ0} is not vanishing, in general. In any case, in supergravity with scalars  duality symmetry was never proven to be  valid in deformed theory on shell.

At this point it is important to mention a progress   towards understanding the role of \E\, symmetry  made by Hillmann
\cite{Hillmann:2009zf}  applying  the ideas of  Henneaux and Teitelboim in \cite{Henneaux:1988gg} to $\cN=8$ supergravity. The price to pay for the manifest \E\, symmetry  in classical $\cN=8$ supergravity was to break manifest four-dimensional general coordinate covariance  of the action, which was a way to produce a `bona fide' \E\,  current and its conservation. In classical theory it was shown in \cite{Hillmann:2009zf} that four-dimensional general coordinate covariance is restored. However, in presence of deformation, the relevant investigation in \cite{Bossard:2011ij} is also restricted to $U(1)$ duality and therefore is also not valid for extended supergravity with scalars.

One possible solution of this issue in BN proposal would be to use an E7 type quartic invariant which does not depend on scalars. This possibility was studied in \cite{Kallosh:2012yy}. We have found, however,  in \cite{Kallosh:2012yy}  that the relevant 4-point vector amplitude is not consistent with supersymmetry.

 We therefore believe that BN proposal \cite{Bossard:2011ij}, claiming that a deformation of classical supergravity is consistent with duality symmetry in $\cN\geq 5$ supergravity, has not yet been substantiated.
 \section{ Noether-Gaillard-Zumino Identity}
Here we bring up some details on Noether-Gaillard-Zumino Identity \cite{Gaillard:1981rj,Aschieri:2008ns,Kallosh:2011dp} in supergravity. This  includes the effect of the deformation in a power series in the deformation parameter $\la$, with examples which give more details on NGZ identity. This Appendix is based on  the work  by Y. Yamada and the author of this paper in the context of our earlier paper \cite{Kallosh:2018mlw}. 

Consider the infinitesimal form of the duality symmetry in eq. \rf{dualrot}
\begin{equation}
\delta \left ( \begin{array}{c}  F\cr  G\cr   \end{array}\right
) \, =\, \left ( \begin{array}{cc} a & b \cr c & d \cr  \end{array} \right )
\left (  \begin{array}{c}   F\cr   G\cr  \end{array}\right ).
\label{infdualrot}
\end{equation}
 We start with an action $S=S(F,\varphi)$. The variation under duality transformation is given by
\begin{align}
\delta S&=\int d^4x \left[\delta \varphi \frac{\delta \cL}{\delta \varphi}+\delta F^+ \frac{\delta \cL}{\delta F^+}+\delta F^- \frac{\delta \cL}{\delta F^-}\right]\nn\\
&=\int d^4x \left[\delta \varphi \frac{\delta \cL}{\delta \varphi}-2i(F^+a+G^+b) G^++2i(F^-a+G^-b) G^-\right],
\end{align}
where $a,b$ are infinitesimal transformation parameter (matrix).\footnote{We omit transpose...etc of $a,b$ in this note and that $a=-d^T$. So $2iF^+aG^+ = iF^+aG^+ - iG^+dF^+$. } According to \cite{Gaillard:1981rj} the vector dependent  variation should take the form
\begin{align}
\delta S=&\int d^4x[ Gb\tilde{G}+Fc\tilde{F}]\nn\\
=&\int d^4x[-iG^+bG^++iG^-bG^--iF^+cF^++iF^-cF^-] \ .
\end{align}
 Equating these two vector dependent expressions gives us the generalized NGZ identity,
\begin{equation}
\int d^4x \left[\delta \varphi \frac{\delta \cL_v}{\delta \varphi}+\left\{-2iF^+aG^+-iG^+b G^++iF^+cF^++{\rm h.c.}\right\}\right]=NGZ_1 + NGZ_2= 0.
\end{equation}
The vector dependent $b$ part of it is presented in eq. \rf{NGZ0}.
If we would assume that $ G= dB$ on shell 
\begin{equation}
(G^{(0)} + \lambda  G^{(1)} + \lambda^2  G^{(2)}+\cdots)_{\mu\nu} \Rightarrow  \partial_\mu B_\nu - \partial_\nu B_\mu  
\label{dB1}\end{equation}
since
\begin{equation}
\partial^\mu (\tilde G^{(0)} + \lambda \tilde G^{(1)} + \lambda^2 \tilde G^{(2)}+\cdots)_{\mu\nu}= 0 \ ,\label{EOM}\end{equation}
on shell, we would conclude that $F\tilde G$ and $G\tilde G$ vanish on shell (not only  $F\tilde F$), so that $NGZ_1$ and  $NGZ_2$ should vanish on shell separately. 

It appears that  $NGZ_1$ and $NGZ_2$ do not vanish separately, they   cancel each other at each order in $\lambda$. But if we would use eq. (\ref {dB1}) they would vanish separately. So, let us present the details of NGZ identity using an action with deformation, as given in \cite{Kallosh:2018mlw}.

Our deformed vector dependent action including scalars is
\begin{equation}
S=\int d^4x \left[-iF^+\cN F^+-iF^+X(\cN-\bar\cN)F^--iF^+X\bar X(\cN-\bar{\cN})F^++{\rm h.c.}\right]
\end{equation}
up to $\mathcal O (\lambda^2)$.
The l.h.s. of the identity at $\mathcal O (\lambda^0)$ is given by
\begin{align}\label{long}
&\int d^4x \left[\delta \cN \frac{\delta \cL_{(0)}}{\delta \cN}+\left\{-2iF^+aG_{(0)}^+-iG_{(0)}^+b G_{(0)}^++iF^+cF^++{\rm h.c.}\right\}\right]\nn\\
=&\int d^4x \left[\delta \cN \frac{\delta \cL_{(0)}}{\delta \cN}+\left\{-2iF^+a\cN F^+-iF^+\cN b \cN F^++iF^+cF^++{\rm h.c.}\right\}\right]\nn\\
=&\int d^4x \left[\delta \cN \frac{\delta \cL_{(0)}}{\delta \cN}+\left\{iF^+(c-2a\cN -\cN b \cN)F^++{\rm h.c.}\right\}\right],
\end{align}
where we have used $G_{(0)}^+=\cN F^+$. Let us compute the first term. The variation of $\cN$ is
\begin{equation}
\delta \cN=c-2\cN a -\cN b\cN,
\end{equation}
and then, the first term becomes
\begin{align}
\delta \cN \frac{\delta \cL_{(0)}}{\delta \cN}=-iF^+(c-2\cN a -\cN b\cN)F^+,
\end{align}
which exactly cancels  other terms in \rf{long}. Thus we confirm the identity at $\mathcal O(\lambda^0)$.

At $\mathcal O(\lambda)$ level, the l.h.s. of the identity is
\begin{align}
&\int d^4x \left[\delta \varphi \frac{\delta \cL_{(1)}}{\delta \varphi}+\left\{-2iF^+aG_{(1)}^+-2iG_{(0)}^+b G_{(1)}^++{\rm h.c.}\right\}\right].
\end{align}
Using $G_{(0)}^+=\cN F^+$ and $G^+_{(1)}=X(\cN-\bar\cN)F^-=-if^{-1}\Delta\bar{f}^{-1}F^-$, we find
\begin{align}
-2iF^+aG_{(1)}^+-2iG_{(0)}^+b G_{(1)}^+=-2F^+(a+\cN b)f^{-1}\Delta\bar f^{-1}F^-,
\end{align}
whereas the first part becomes
\begin{align}
\delta \varphi \frac{\delta \cL_{(1)}}{\delta \varphi}=&-2F^+\delta f^{-1}\Delta\bar f^{-1}F^--2F^+f^{-1}\Delta\delta\bar f^{-1}F^-\nn\\
=&+2F^+(a+\cN b) f^{-1}\Delta\bar f^{-1}F^-+2F^+f^{-1}\Delta\bar f^{-1}(a+b\bar\cN)F^-\nn\\
=&2F^+(a+\cN b) f^{-1}\Delta\bar f^{-1}F^-+{\rm h.c.}.
\end{align}
Here we have used the identity $\delta f^{-1}=\delta f^{-1}f f^{-1}=-f^{-1}\delta f f^{-1}=-(a+\cN b)f^{-1}$. Hence the terms cancel to each other, and the identity holds.

The second order part is more involved. Let us compute the part $\delta \varphi \frac{\delta \cL_{(2)}}{\delta \varphi}$. The second order part of the Lagrangian is
\begin{align}
\cL_{(2)}=&-iF^+X\bar X(\cN-\bar \cN)F^++{\rm h.c.}\nn\\
=&-F^+f^{-1}\Delta f\bar f^{-1}\bar\Delta f^{-1}F^++{\rm h.c.}.
\end{align}
The variation of the Lagrangian is
\begin{align}
\delta\cL_{(2)}=&-F^+\delta f^{-1}\Delta f\bar f^{-1}\bar\Delta f^{-1}F^+-F^+f^{-1}\Delta \delta f\bar f^{-1}\bar\Delta f^{-1}F^+-F^+f^{-1}\Delta f\delta\bar f^{-1}\bar\Delta f^{-1}F^+\nn\\
&-F^+f^{-1}\Delta f\bar f^{-1}\bar\Delta \delta f^{-1}F^++{\rm h.c.}\nn\\
=&F^+(a+\cN b)f^{-1}\Delta f\bar f^{-1}\bar\Delta f^{-1}F^+-F^+f^{-1}\Delta f(a+\cN b)\bar f^{-1}\bar \Delta f^{-1} F^+\nn\\
&+F^+f^{-1}\Delta f (a+\bar\cN b)\bar f^{-1}\bar\Delta f^{-1}F^++F^+f^{-1}\Delta f\bar f^{-1}\bar\Delta f^{-1}(a+b\cN) F^++{\rm h.c.}\nn\\
=&2F^+(a+\cN b)f^{-1}\Delta f\bar f^{-1}\bar\Delta f^{-1}F^+-F^+f^{-1}\Delta f (\cN-\bar\cN)b\bar f^{-1}\bar\Delta f^{-1}F^++{\rm h.c.}\nn\\
=&2F^+(a+\cN b)f^{-1}\Delta f\bar f^{-1}\bar\Delta f^{-1}F^++iF^+f^{-1}\Delta \bar f^{-1}b\bar f^{-1}\bar\Delta f^{-1}F^++{\rm h.c.}\nn\\
=&+2iF^+a G^+_{(2)}+2iG^+_{(0)}bG^+_{(2)}-iG^-_{(1)}bG^-_{(1)}+{\rm h.c.},
\end{align}
where we have used the identity $\delta f^{-1}=-(a+\cN b)f^{-1}$, $G_{(2)}^+=X\bar X(\cN-\bar \cN)F^+=-if^{-1}\Delta f\bar f ^{-1}\bar\Delta f^{-1}F^+$ and $G^-_{(1)}=i\bar f^{-1}\bar\Delta f^{-1}F^+$. The rest part of the l.h.s. of the identity is exactly the same form with over all minus sign. Hence, we have confirmed the identity at the second order of $\lambda$.

As suggested in Appendix E, the fact that the consistency of the BN proposal in presence of scalars remains unproven, is not accidental. Whereas in absence of scalars NGZ identity easily supports the consistency of the deformation, like in $U(1)$ duality examples, NGZ identity presents an obstruction to BN proposal  in $\cN\geq 5$ supergravity with scalars.

\bibliographystyle{JHEP}
\bibliography{refs}

\end{document}